\documentclass[twocolumn]{aastex62}
\usepackage{amsmath}
\usepackage{amsfonts}

\usepackage[normalem]{ulem}

\allowdisplaybreaks
\graphicspath{{./}{figures/}}

\received{March 7, 2022}
\revised{May 17, 2022}
\accepted{May 18, 2022}

\submitjournal{ApJL}

\shorttitle{FRBs from rapid reconnection}
\shortauthors{Mahlmann et al.}

\begin{document}

\title{Electromagnetic fireworks: Fast radio bursts from rapid reconnection in the compressed magnetar wind}

\correspondingauthor{J. F. Mahlmann}
\email{mahlmann@princeton.edu}

\author[0000-0002-5349-7116]{J. F. Mahlmann}
\affil{Department of Astrophysical Sciences, Peyton Hall, Princeton University, Princeton, NJ 08544, USA}

\author[0000-0001-7801-0362]{A. A. Philippov}
\affiliation{Center for Computational Astrophysics, Flatiron Institute, New York, NY 10010, USA}
\affil{Department of Physics, University of Maryland, College Park, MD 20742, USA}

\author[0000-0001-7572-4060]{A. Levinson}
\affiliation{The Raymond and Beverly Sackler School of Physics and Astronomy, Tel Aviv University, Tel Aviv 69978, Israel}

\author[0000-0001-9179-9054]{A. Spitkovsky}
\affiliation{Department of Astrophysical Sciences, Peyton Hall, Princeton University, Princeton, NJ 08544, USA}

\author[0000-0001-8939-6862]{H. Hakobyan}
\affiliation{Computational Sciences Department, Princeton Plasma Physics Laboratory (PPPL), Princeton, NJ 08540, USA}
\affiliation{Physics Department \& Columbia Astrophysics Laboratory, Columbia University, New York, NY 10027, USA}

\begin{abstract}

One scenario for the generation of fast radio bursts (FRBs) is magnetic reconnection in a current sheet of the magnetar wind. Compressed by a strong magnetic pulse induced by a magnetar flare, the current sheet fragments into a self-similar chain of magnetic islands. Time-dependent plasma currents at their interfaces produce coherent radiation during their hierarchical coalescence. We investigate this scenario using 2D radiative relativistic particle-in-cell simulations to compute the efficiency of the coherent emission and to obtain frequency scalings. Consistent with expectations, a fraction of the reconnected magnetic field energy, $f\sim 0.002$, is converted to packets of high-frequency fast magnetosonic waves which can escape from the magnetar wind as radio emission. In agreement with analytical estimates, we find that magnetic pulses of $10^{47}\text{erg}\;\text{s}^{-1}$ can trigger relatively narrowband GHz emission with luminosities of approximately $10^{42}\text{erg}\;\text{s}^{-1}$, sufficient to explain bright extragalactic FRBs. The mechanism provides a natural explanation for a downward frequency drift of burst signals, as well as the $\sim 100\;\text{ns}$ substructure recently detected in \object{FRB 20200120E}.
\end{abstract}

\submitjournal{ApJ}


\section{Introduction}
\label{sec:Introduction}

Fast radio bursts (FRBs) are extremely bright, short-duration ($\lesssim$ few ms) radio pulses observed in the frequency range $0.1-10$GHz \citep[][]{Chime2019b} at inferred cosmological distances. They are typically highly polarized, highly dispersed, and exhibit
complex temporal and spectral structures (likely caused, at least in part, by propagation effects). 
The enormous brightness temperatures that characterize FRBs are indicative of coherent radio emission. Among the about $10^3$ FRBs that were already recorded, tens are known to repeat \citep{Spitler2014,Spitler2016,Chime2019,Chime2019a}, 
and one of these shows evidence for a $\sim 16$ day periodicity \citep{chime2020b}, disfavoring cataclysmic origin (e.g., mergers of compact objects). Despite this observational track record, the nature of FRBs remains unknown, and currently proposed models \citep[see][]{Lyubarsky2021} await consolidation.  The recent association of a (faint) FRB with the Galactic magnetar SGR 1935+2154, featuring two radio bursts accompanied by much more powerful X-ray bursts \citep{Chime2020,Bochenek2020,Kirsten2020,Lin2020,Mereghetti2020,Scholz2020,Telegram2020,Israel2021,Li2021,Ridnaia2021}, has lent support to the hypothesis that at least some FRBs (notably the class of repeaters) are produced by magnetars. In this paper, we adopt this view.

Magnetar FRB models are broadly divided into two main types; magnetospheric models, in which the waves are generated close to the star \citep[e.g.,][]{lyutikov2020}, and so-called `far away' models in which the emission is produced at larger radii. Among the latter type,
two compelling magnetar-powered FRB mechanisms were developed alongside the ongoing observational surveys. One is the \emph{shock-mediated} synchrotron maser model, which associates ring-like distributions of electrons gyrating in the ordered magnetic field of the shock with the emission of coherent electromagnetic (EM) waves \citep{Lyubarsky2014,Ghisellini2016,Beloborodov2017,Waxman2017,Gruzinov2019,Metzger2019,Plotnikov2019,Beloborodov2020,Margalit2020}. In the other, a \emph{reconnection-mediated} fast magnetosonic (FMS) wave generation in the outer magnetosphere emerges as the natural consequence of merging plasmoids in a compressed current sheet \citep{Lyubarsky2020}. In this scenario, a magnetospheric current sheet is compressed by a magnetic low-frequency pulse (LFP) emerging close to the magnetar due to a flare. The thereby triggered reconnection episode produces high-frequency EM waves during hierarchical plasmoid mergers.\footnote{Fast-wave emission by plasmoid mergers is a universal mechanism that does not require strong field compression. A reconnecting current sheet in the magnetar wind can power FMS waves by the same mechanism. However, the frequency of this emission is too low to be observed in the GHz range. On the other hand, the frequency emitted by plasmoids merging in a current sheet produced in a flare close to the magnetar is too high \citep{Lyubarsky2020,Most2020}.}

The hypothesis of FRB generation by \textit{reconnection-mediated} FMS wave production is the subject of this paper. We supplement the theoretical framework and application to the magnetar magnetosphere laid out in \citet{Lyubarsky2020,Lyubarsky2021} by demonstrating its validity in numerical experiments, namely, large-scale relativistic particle-in-cell (PIC) simulations. To this end, we build upon previous studies explaining radio nanoshots in the context of the Crab pulsar via FMS wave generation by merging plasmoid structures in a reconnection layer \citep{Lyubarsky2019,Philippov2019}. Radiative losses were not taken into account in the first numerical study of the reconnection-mediated coherent emission mechanism by \citet{Philippov2019}. In the magnetar magnetosphere, the reconnection heated plasma in the compressed layer of the LFP-current sheet interaction is, however, subject to strong synchrotron cooling. We explore these effects quantitatively in the numerically accessible regime of low to intermediate cooling losses. Interestingly, one recent FRB observation identifies pulse substructures that are similar to those found in the Crab \citep[\object{FRB 20200120E}, see][]{Majid2021}. Such a temporal structure is not expected in far away maser models. Below, we show that it can be produced naturally during reconnection-mediated FRB generation.

This paper is organized as follows. Section~\ref{sec:PlasmaScaling} reviews the relevant scalings of magnetospheric plasma and LFP properties, as well as the frequency dependence of the outgoing FMS waves. Section~\ref{sec:Simulations} presents the numerical setup (Section~\ref{sec:SimulationSetup}) and simulation results (Section~\ref{sec:result}). We examine the general dynamics of the interaction between an LFP and a current sheet in Section~\ref{sec:largesystems}. Section~\ref{sec:amplitudedependence} reviews the imprint of different amplitudes of the LFP on the outgoing FMS waves. We study the effect of synchrotron cooling in Section~\ref{sec:cooling}. Finally, we discuss several astrophysical considerations in Section~\ref{sec:dicsussion} and summarize in Section~\ref{sec:conclusion}.

\section{Plasma environment and scalings}
\label{sec:PlasmaScaling}

\begin{figure}
  \centering
  \includegraphics[width=0.47\textwidth]{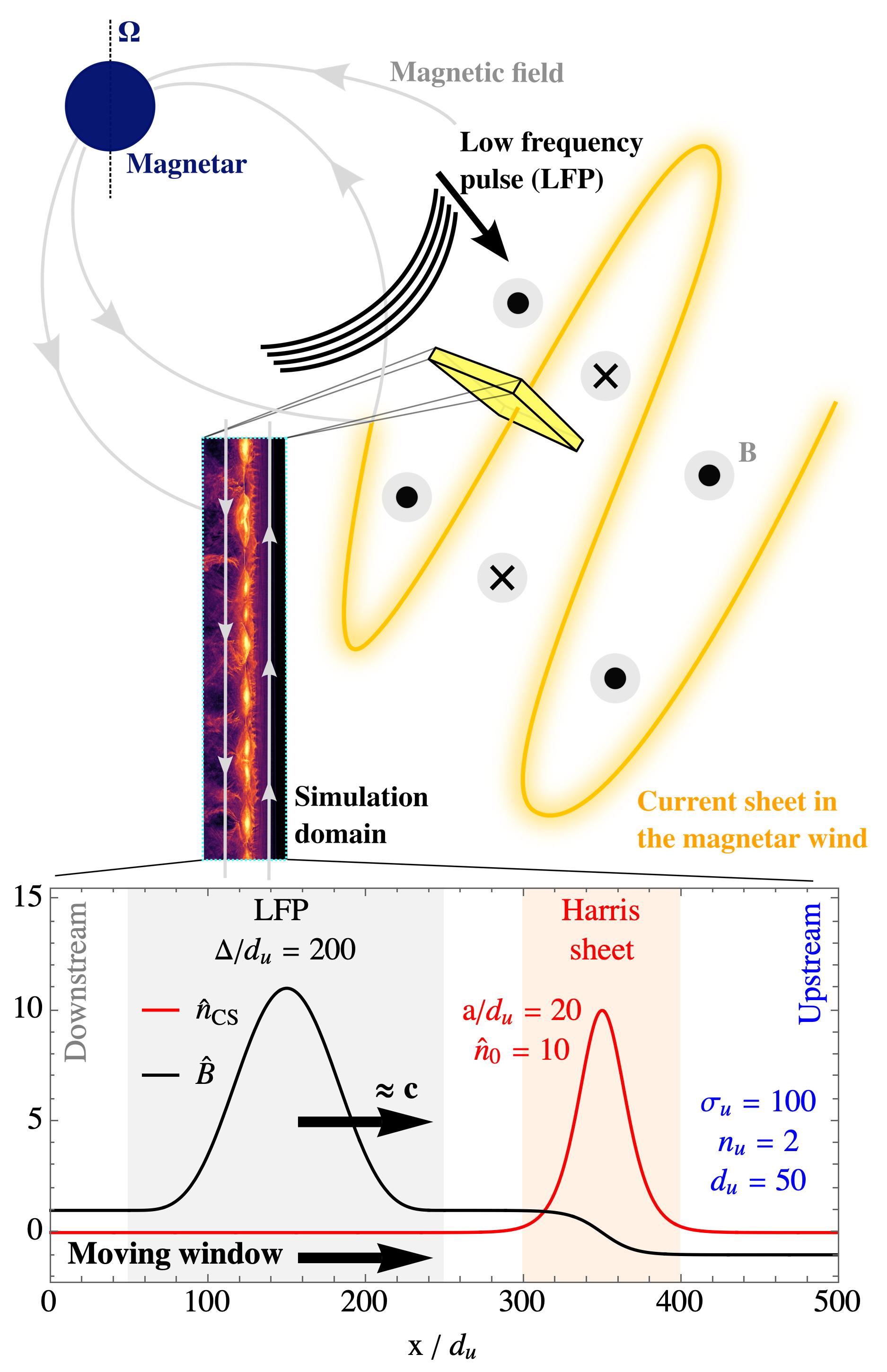}
  \vspace{-18pt}
  \caption{Schematic impression of the simulation setup (bottom) and its global astrophysical context (top). The simulation domain mimics a segment of the striped magnetar wind, located in the vicinity of the light cylinder. An LFP emerging close to the magnetar propagates outwards and collides with a Harris-type current sheet of the striped wind. We model their subsequent interaction in a moving window, tracking the propagation of the fast pulse at the speed of light through a system of infinite length (and energy supply).}
\label{fig:simsetup}
\end{figure}

The proposed FRB generation mechanism happens close to the magnetar light cylinder. Figure~\ref{fig:simsetup} gives a cartoon impression of its astrophysical context and simulation setup. A magnetar with nonaligned rotation and magnetic axes generates a striped wind separated by a Harris-type current sheet (yellow). An LFP emerging close to the magnetar (induced by a magnetar flare) travels outwards compresses the magnetic fields and eventually the current sheet in the magnetar wind (with an amplitude relative to the background magnetic field of $B/B_0\gtrsim 10^4$, see Section~\ref{sec:dicsussion}). We model the reconnection in such a compressed current sheet that is pushed by a strong magnetic LFP traveling at relativistic speeds. For the remainder of this paper, the index `u' will denote quantities in the upstream, or in the magnetar wind in front of the LFP. The index `p' stands for quantities associated with the LFP, and a prime indicates quantities in the plasma rest frame. To understand the different processes at play, we review relevant plasma properties on a fundamental level in this introductory section \citep[see][]{Lyubarsky2020}. Specifically, we answer several questions that outline the expected length and time scales and their imprint on the spectrum of outgoing FMS waves.

First, we review how much separation to expect between the plasma scales and the pulse width in the astrophysical context. This work examines reconnection processes happening in the outer magnetar magnetosphere, close to its light cylinder \citep{Lyubarsky2020}. The multiplicity of a pair plasma is commonly defined as $\mathcal{M}=|e|\left(n_++n_-\right)c/j$ \citep[see][]{Beloborodov2013,Chen2017}. Thus, the pair density becomes $n=\mathcal{M}j/(ec)$, where $j$ is an appropriate current density. The Goldreich-Julian charge density \citep{Goldreich_Julian_1969ApJ...157..869}, $j_{\rm GJ}=\rho_{\rm GJ} c=-\mathbf{\Omega}\cdot\mathbf{B}/(2\pi)$ for a dipolar magnetosphere with field line angular velocity $\mathbf{\Omega}$ can be used as a proxy, yielding the plasma skin depth:
\begin{align}
    d_{\rm u}=\frac{c}{\omega_{\rm pu}}=\left(\frac{\gamma_{\rm u}m_e c^2}{4\pi n_u e^2}\right)^{1/2}\approx R_{\rm LC}^2\left(\frac{\gamma_{\rm u}m_e c^2}{2\mathcal{M} e \mu}\right)^{1/2}.
    \label{eq:SkinDepth}
\end{align}
Here, $\omega_{\rm pu}$ is the relativistic electron-positron plasma frequency, $c$ is the speed of light, $m_e$ is the electron mass, $\gamma_{\rm u}$ the bulk Lorentz factor of the upstream flow, and $n_{\rm u}$ is the pair density. In the approximation of Equation~(\ref{eq:SkinDepth}), we evaluated $j_{\rm GJ}$ at the light cylinder, $R_{\rm LC}=c/\Omega=cP/2\pi$, by assuming a dipole magnetic field along the equatorial plane, namely $B\approx \mu / r^3$. Here, $\mu=B_* R_*^3$ denotes the magnetic moment of the dipole field, evaluating to $\mu\approx 10^{33}\text{erg}\;\text{G}^{-1}$ for common choices of the stellar radius, $R_*=10\;\text{km}$, and the stellar magnetic field, $B_*= 10^{15}\text{G}$. We can constrain expression (\ref{eq:SkinDepth}) to
\begin{align}
\begin{split}
    d_{\rm u}\approx&\; 2.1\times 10^3\gamma_{\rm u}^{1/2}\left(\frac{100}{\mathcal{M}}\right)^{1/2}\left(\frac{10^{15}\text{G}}{B_*}\right)^{1/2}\left(\frac{P}{1\text{s}}\right)^2\text{cm}\label{eq:duscaling}.
\end{split}
\end{align}
With this plasma scale, we can define a dimensionless pulse width $\Delta/d_{\rm u}$. The typical pulse length $\Delta=\tau c$ is related to a typical FRB duration $\tau$ \citep{Lyubarsky2020}, such that common values of $\Delta/d_{\rm u}$ can be estimated as
\begin{align}
    \Delta/d_{\rm u}\approx 1.4\times 10^4\;\gamma_{\rm u}^{-1/2}\left(\frac{\tau}{1\text{ms}}\right).
\end{align}
This is a macroscopic width with significant scale separation to the microscopic reconnection processes that take place across a few (compressed) plasma skin depths. Currently available techniques and resources allow us to model pulses with the width of a small fraction of this estimate (as shown in Section~\ref{sec:Simulations}).

Then, we can also establish the relevant time scales for the LFP-current sheet interaction. A strong magnetic pulse propagates through a relativistic pair plasma with the magnetization $\sigma_{\rm u}\gg 1$. In a steady state, an LFP moving across the background magnetic field $B_{\rm u}$ with the speed of light can be characterized (in the upstream frame) by the continuity condition $\hat{n}_{\rm p}\equiv n_{\rm p}/n_{\rm u}=B_{\rm p}/B_{\rm u}\equiv\hat{B}_{\rm p}$, as well as (see Appendix~\ref{sec:appendixA})
\begin{align}
   \gamma_{\rm p}=\hat{B}_{\rm p}\left(2\hat{B}_{\rm p}-1\right)^{-1/2}.
   \label{eq:pulsefactors}
\end{align}
While the LFP propagates with a characteristic rest-frame velocity, $s_{\rm fast}$, the supporting plasma drifts with $\gamma_{\rm p}$ relative to the upstream frame (which streams at $\gamma_u$ with respect to the magnetar frame) and $\Gamma_{\rm p}$ relative to the distant observer (with realistic values of $\Gamma_{\rm p}\sim 100$). We obtain the interaction time $t_{\rm i}$ from $t'_{\rm i}=\Delta'/s_{\rm fast}$, with pulse width $\Delta$. In the limit of $\hat{B}_{\rm p}\gg 1$
\begin{align}
\begin{split}
        t_{\rm i}\propto \Gamma_{\rm p}^2 \frac{\Delta}{s_{\rm fast}}\approx \frac{1}{2}\frac{\Delta}{c}\hat{B}_{\rm p},
\end{split}
    \label{eq:interaction}
\end{align}
in the magnetar frame. In the last approximation and throughout the presented simulations (conducted in the frame of the magnetar wind), we use $\gamma_{\rm u}=1$. We confirm the scaling of Equation~(\ref{eq:interaction}) with the characteristic pulse crossing time $t_{\rm p}=\Delta/c$ in Appendix~\ref{sec:appendixA1}. We note that $t_{\rm i}$ is the relevant scale for the interaction dynamics; in an \emph{infinite} system, it determines the amount of magnetic flux of polarity opposite to the one of the pulse, $\phi_{\rm u}\sim c t_{\rm i} B_{\rm u}L_z$, processed from the upstream. The magnetic flux initially carried by the LFP can be found as $\phi_{\rm p}\sim c t_{\rm p} B_{\rm p} L_z $. Thus, using Equation~(\ref{eq:interaction}), one finds that for the reconnection to occur during the whole duration $t_{\rm i}$, an amount of opposite flux in the upstream should be comparable to the flux in the pulse, $\phi_{\rm u}\sim \phi_{\rm p}/2$. In this case, the observed signal duration is of order $t_{\rm p}$ (see simulation results in Section~\ref{sec:result}). If reconnection ceases because of the limited upstream flux, $\phi_{\rm u}\ll\phi_{\rm p}/2$, the radiation of FMS waves stops operating and the signal duration can be shorter than $t_{\rm p}$ (see Section~\ref{sec:dicsussion}).

Next, we consider \textit{how much energy can be extracted from the LFP by reconnection}. The FMS pulse compresses the current sheet and enhances the magnetic field at the tangential discontinuity by a factor of $\hat{B}_{\rm p}$. The energy converted by reconnection can be determined by evaluating the appropriate\footnote{The Poynting flux flows into the current sheet from two directions, but only half of the energy is dissipated by reconnection \citep{Sironi2015}. These effects effectively cancel each other out in this approximate theoretical estimate.} Poynting flux in the rest frame of the plasma:
\begin{align}
    \text{d}e'_{\rm r}=\frac{L_y L_z}{4\pi}E'_{\rm p}B'_{\rm p} \text{d}t'= \frac{L_yL_z}{4\pi} c\beta_{\rm rec} \left(B'_{\rm p}\right)^2 \text{d}t'.
    \label{eq:energyintegand}
\end{align}
We used $E'_{\rm p}=c\beta_{\rm rec}B'_{\rm p}$ with reconnection rate $\beta_{\rm rec}$, layer length $L_y$, and layer depth $L_z$. \footnote{We assume that $\beta_{\rm rec}$ does not vary significantly for the duration of the interaction. This assumption can be justified as follows. The extent of causally connected regions along the current sheet is limited by $L'_x\times c/s_{\rm fast}\approx L'_x$. During the LFP-current sheet interaction, the growing plasmoid structures at most cover a fraction of the characteristic length scale $L'_x$ \citep{Sironi2016}. Thus, plasmoids do not become large enough to dominate the current layer and to reduce the rate of reconnection as long as $1\lesssim L'_y/L'_x\approx L_y/(L_x \gamma_{\rm p})$. Throughout this study, we simulate system sizes with $L'_y/L'_x\sim 0.35 - 1.75$ and do not find a notable effect on the interaction dynamics. In the realistic magnetar wind, the system's aspect ratio is likely even larger, $L'_y/L'_x\approx R_{\rm LC} / (\Gamma_{\rm p}\tau c) \sim 3.3$ for the expected pulse amplitudes (see Section~\ref{sec:dicsussion}). The limiting case of this argument is $L'_y/L'_x\rightarrow 0$, in which the reconnection rate will notably decrease. We examine the case of finite $L'_x$ and $L'_y\rightarrow 0$ for a 1D setup in Appendix~\ref{sec:appendixA1}, where reconnection does not occur and no energy is dissipated during the interaction.} For an \emph{infinite} system size of length $L_y$ we can use Equation~(\ref{eq:interaction}) to write
\begin{align}
    e'_{\rm r} = \frac{L_y L_z}{4\pi} \beta_{\rm rec} \left(B'_{\rm p}\right)^2 \Delta'.
    \label{eq:recenergy}
\end{align}
Normalized to $e'_{\rm p}=\Delta' L_y L_z\left(B'_{\rm p}\right)^2 /(8\pi)$, the initial pulse energy, one obtains $e_{\rm r}/e_{\rm p}\approx 2\beta_{\rm rec}$. In contrast, for a \emph{finite} system size, \citet{Lyubarsky2020} estimates the relevant area enclosing the reconnecting magnetic flux by the scale that is set by the light cylinder radius $R_{\rm LC}$. Combined with the magnetic field strength at the light cylinder, $B_{\rm LC}$, direct integration of Equation~(\ref{eq:energyintegand}) yields
\begin{align}
    e_{\rm r}\propto\hat{B}_{\rm p} B_{\rm LC}^2 R_{\rm LC}^3,
    \label{eq:erfinite}
\end{align}
reproducing the estimate by \citet[][Equation 5]{Lyubarsky2020}. For the models considered throughout this work, we focus on the somewhat simplified \emph{infinite} system sizes.

To establish the relevance of the considered process for astronomical observations we stress how plasmoid mergers during reconnection translate into an outgoing coherent emission and how (potentially strong) synchrotron cooling can affect its peak frequency. The characteristic frequency of the outgoing waves is given by \citep[see][]{Lyubarsky2020}
\begin{align}
    \omega'=\frac{c}{\xi a'}\; ,
\end{align}
where $\xi$ is the ratio between the size of the plasmoids relative to the thickness, $a'$, of the \emph{compressed} current sheet, commonly approximated by $\xi\sim 10-100$. The layer thickness is the main ingredient in determining the magnetar frame frequency, $\nu$, of the outgoing FMS waves, $\nu\approx\Gamma_{\rm p} \omega'/(2\pi)=\Gamma_{\rm p} c / (2\pi \xi a')$. One can estimate the layer width $a'$ to be comparable to the typical Larmor radius of accelerated particles in the layer, a common choice in studies of collisionless reconnection \citep[e.g.,][]{Uzdensky2013, Uzdensky2016}. We, thus, assume $a'=\zeta\rho'_{\rm L}$ with $\zeta\sim 1-10$ and $\rho'_{\rm L}=\rho'\left\langle\gamma'\right\rangle=m_e c^2/(e B')\left\langle\gamma'\right\rangle$ being the Larmor radius in the rest frame of the plasma, where $\rho'=m_e c^2/(e B')$ is the nominal Larmor radius. It can be expressed in terms of the gyrofrequency $\omega'_B=e B'/(m_e c)$ as $\rho'_{\rm L}=\left\langle\gamma'\right\rangle c/\omega'_B$. The outgoing wave frequency, thus, is
\begin{align}
    \nu=\frac{1}{2\pi\xi\zeta}\frac{\omega_B}{\left\langle\gamma'\right\rangle}.
    \label{eq:outfrequency}
\end{align}
In other words, Equation~(\ref{eq:outfrequency}) describes how the \emph{pulse amplitude} and \emph{synchrotron cooling} strength affect the (peak) frequency of the outgoing FMS wave signature. Assuming a balance between synchrotron cooling and heating in the reconnection layer, the radiation-reaction limit $\left\langle\gamma'\right\rangle\approx\gamma'_{\rm rad}$ \citep[][]{Uzdensky2013,Hakobyan2019}, Equation~(\ref{eq:outfrequency})  becomes 
\begin{align}
    \nu=\frac{1}{2\pi\xi\zeta}\left(\frac{2}{3}\frac{r_e}{c\beta_{\rm rec} \Gamma_{\rm p}}\right)^{1/2}\omega_B^{3/2}
    \label{eq:outfrequencylab}
\end{align}
in the compression zone, which is in good agreement with Equation~(12) of \citet{Lyubarsky2020}. The limit of slow cooling is approximated by $\left\langle\gamma'\right\rangle\approx\sigma'=\sigma_{\rm u} \hat{B}_{\rm p}/\Gamma_{\rm p}$, such that Equation~(\ref{eq:outfrequency}) becomes
\begin{align}
\nu=\frac{1}{2\pi\xi\zeta}\frac{c}{\rho_{\rm L u}}\Gamma_{\rm p}\propto \hat{B}_{\rm p}^{1/2}.
\label{eq:outgoinguncooled}
\end{align}

Finally, we ask if the strongly compressed reconnection layer finds sufficient plasma supply during the interaction. We quantify this supply by the charge starvation coefficient (calculated in the plasma rest frame):
\begin{align}
    \alpha' = \frac{1}{4\pi}\frac{|\nabla'\times\delta\mathbf{B}'|}{n'e}\approx\frac{B'}{4\pi n' ea'}.
    \label{eq:alphaparameter}
\end{align}
The pressure balance in the current sheet requires
\begin{align}
    n'_{\rm cs} m_e c^2 \left\langle\gamma'\right\rangle = \frac{B'^2}{8\pi}.
\end{align}
With our previous definition of the reconnection layer thickness in terms of the Larmor radius (see above Equation~\ref{eq:outfrequency}) we, thus, obtain
\begin{align}
\begin{split}
    \alpha' &< \frac{B'}{4\pi e} \frac{8\pi m_e c^2 \left\langle\gamma'\right\rangle}{B'^2}\frac{e B'}{\zeta m_e c^2 \left\langle\gamma'\right\rangle}=\frac{2}{\zeta}.
\label{eq:alphasecond}
\end{split}
\end{align}
With typical values of $\zeta\sim$ a few, $\alpha'<1$, such that charge starvation is not dominant throughout the interaction. In the upstream frame, we expect the layer width to scale as $\hat{B}_{\rm p}^{-1/2}$. The assumption $a'=\zeta\rho'_{\rm L}>\rho'_{\rm u}\sigma'_{\rm u}$ is verified in Appendix \ref{sec:appendixA1}, where we find $\zeta\approx 2$.

\section{Simulations}
\label{sec:Simulations}

We conduct 2D PIC simulations of a relativistic pair plasma in a current sheet compressed by a magnetic pulse with the \textsc{Tristan-MP v2} code \citep{tristanv2}. 

\subsection{Simulation setup}
\label{sec:SimulationSetup}

We initialize a strong LFP and a Harris-type current sheet of width $a$ (see Figure~\ref{fig:simsetup}) according to the kinetic equilibrium configurations used, e.g., in \citet[][]{Kirk2003}. The background magnetic field $\mathbf{B}_{\rm u}=B_{\rm cs}\mathbf{\hat{y}}$ and density profile $n(x)=n_{\rm u}\left(1+n_{\rm cs}\right)$ are
\begin{align}
    B_{\rm cs}(x)&=B_{\rm u}\;\tanh\left(\frac{x-x_0}{a}\right)\label{eq:BCS},\\
        n_{\rm cs}&=\hat{n}_{0}\;\text{sech}^2\left(\frac{x-x_0}{a}\right)\label{eq:nCS},
\end{align}
where we choose the upstream magnetization $\sigma_{\rm u}=100$. Here, $\hat{n}_{0}$ is the ratio between the current sheet and upstream particle density. Our default choices of $\hat{n}_{0}=10$ and $a/d_{\rm u}=20$ stabilize the current sheet so that in isolation it would not develop significant reconnection or tearing modes for the duration of the simulations. The pulse of width $\Delta/d_{\rm u}=200$ is initially located at $x_{\rm p}$. We choose its magnetic field in the interval $x\in\left[x_{\rm p}-\Delta/2,x_{\rm p}+\Delta/2\right]$ as the smooth profile
\begin{align}
    B_{\rm p}(x)=\mathcal{A}\cos\left[\frac{\pi\left(x-x_{\rm p}\right)}{\Delta}\right]^4,
    \label{eq:pulseprofile}
\end{align}
such that $B_y = B_{\rm u}+B_{\rm p}$, and $E_z = -B_{\rm p}$. The dimensionless amplitude factor $\mathcal{A}=\hat{B}_{\rm p}(x_{\rm p})=B_{\rm p}(x_{\rm p})/B_{\rm u}(x_{\rm p})$ denotes the maximum compression. In the high magnetization limit, the LFP moves with a velocity close to the speed of light (Appendix~\ref{sec:appendixA}). To avoid prohibitively large simulation domains along the propagation direction, we employ moving window boundary conditions \citep{Bruhwiler2001}. Specifically, we replenish the upstream environment while shifting the domain along the $x$-direction with the speed of light.

The continuity equation implies the plasma density scaling as $n_{\rm p}=n_{\rm u}\hat{B}_{\rm p}$. To resolve the plasma skin depth and, equally, reduce numerically induced fluctuations in the highly nonlinear pulse, we choose a background skin depth of $d_{\rm u}=50$ cells. The skin depth throughout the pulse then varies as $\hat{d}_{\rm p}=(\hat{\gamma}_{\rm p}/\hat{n}_{\rm p})^{1/2}$. When probing pulse amplitudes in the range of $\mathcal{A}\in\left[5,10,20,30\right]$, the skin depth of the isolated pulse is approximately $d_{\rm p}\in\left[29,23,20,18\right]$ cells. The chosen range of amplitudes allows us to study the scalings of different physical processes (Section~\ref{sec:PlasmaScaling}), while the realistic pulse amplitudes of $\mathcal{A}\gtrsim 10^4$ are numerically prohibitive. In the evaluation of reconnection in a current layer of length $L_y$, the effective length given by the ratio $\kappa\equiv L_y/\rho'\sigma' = L_y/\rho_{\rm u}\sigma_{\rm u}$ is the dominant parameter that determines the properties of the reconnection layer \citep{Werner2015,Sironi2016}. Specifically, if $\kappa\gtrsim 100$, the mechanisms of particle acceleration are not affected by the system size, and plasmoid chains are abundantly produced. The 2D simulation domain extends across the grid spanned by $L_x\times L_y=500\; d_{\rm u}\times d_{\rm u}\sigma^{1/2}_{\rm u}\kappa =25000\times 500\kappa$ cells, where the parameter $\kappa\in\left[30,150\right]$ varies throughout our simulations to ensure convergence of FMS wave spectra and power. In the upstream, we initialize a fixed number of PIC particles per cell ($n_{\rm ppc, u}$). While we checked the convergence of our results for a $n_{\rm ppc, u}\in\left[1,2,5\right]$ per species, we choose the intermediate value of $n_{\rm ppc,u}=2$ for our parameter space explorations.

Several techniques are employed to accommodate the plasma dynamics emerging during the interaction of a strong EM pulse with a current sheet. We reduce (numerical) dispersion errors in the finite-difference time-domain method by using optimized stencils in the EM field solver in combination with a beneficial choice of time step \citep[fixing the Courant number to $0.5$, see][]{Blinne2018}. Vay's optimized particle pusher \citep{Vay2008} further improves capturing the relativistic $\mathbf{E}\times\mathbf{B}$ drift motion induced by the strong LFP. By studying the propagation of a single nonlinear LFP, we found that relatively high resolutions are required to suppress numerical instabilities. Additionally, we use $64$ filter passes for current deposition. To accommodate large density imbalances in some of the presented simulations we use a current-conserving particle merging algorithm in high-density regions (Appendix~\ref{sec:appendixB}). 

\subsection{Results}
\label{sec:result}

\begin{figure*}
  \centering
  \includegraphics[width=0.98\textwidth]{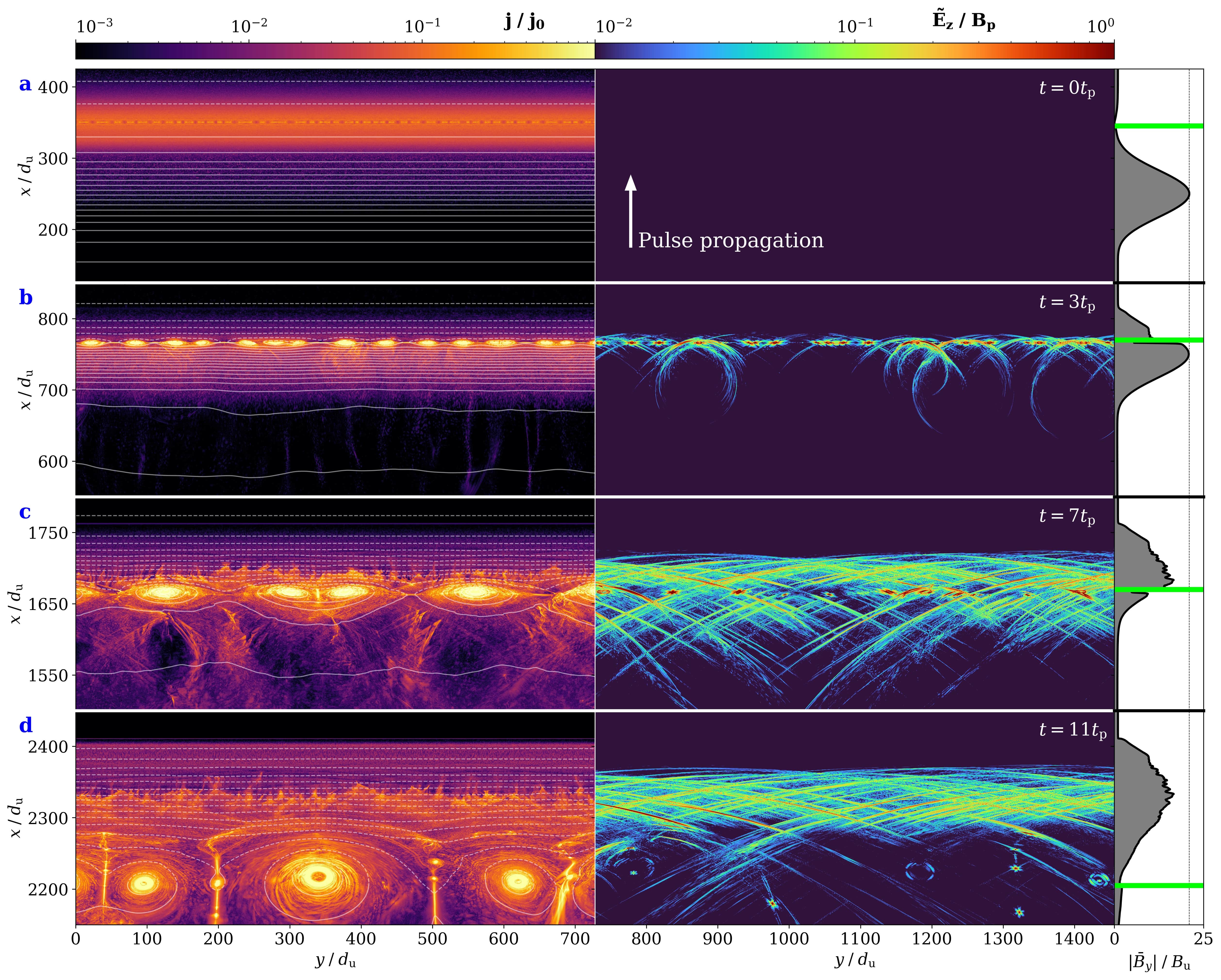}
\vspace{-8pt}
  \caption{Different phases of the interaction between an LFP ($\mathcal{A}=20$) and a current sheet in a large system ($\kappa\approx 150$). The pulse propagates upwards into the upstream. We display - along the full extent $L_y$ with the $y$-direction as the bottom axis - the current density with semitransparent magnetic equipotentials (left panels). Dashed and solid potential surfaces indicate the direction of the magnetic field $B_y$. The middle panels show the out-of-plane electric field component of the emerging FMS waves. We display four stages of the interaction: (a) LFP and noncompressed current sheet previously to their interaction, ({b}) compression of the current sheet and onset of reconnection, ({c}) large plasmoid mergers during maximum compression on top of the pulse, and ({d}) decoupling with secondary current sheets in the wake of the LFP pulse. The right panels display $y$-averaged profiles of the pulse magnetic field and the location of the current sheet (green line). An animation tracking the
pulse propagating upwards into the upstream of a moving domain is available in the online version of this article and as supplemental material \citep{SupplementaryMediaA}.}
\label{fig:largesystems}
\end{figure*}

This section presents various aspects of the proposed mechanism that effectively transforms a fraction of the reconnected energy into high-frequency coherent FMS waves during plasmoid mergers. 

\subsubsection{Interaction dynamics}
\label{sec:largesystems}

\begin{figure}
  \centering
  \includegraphics[width=0.47\textwidth]{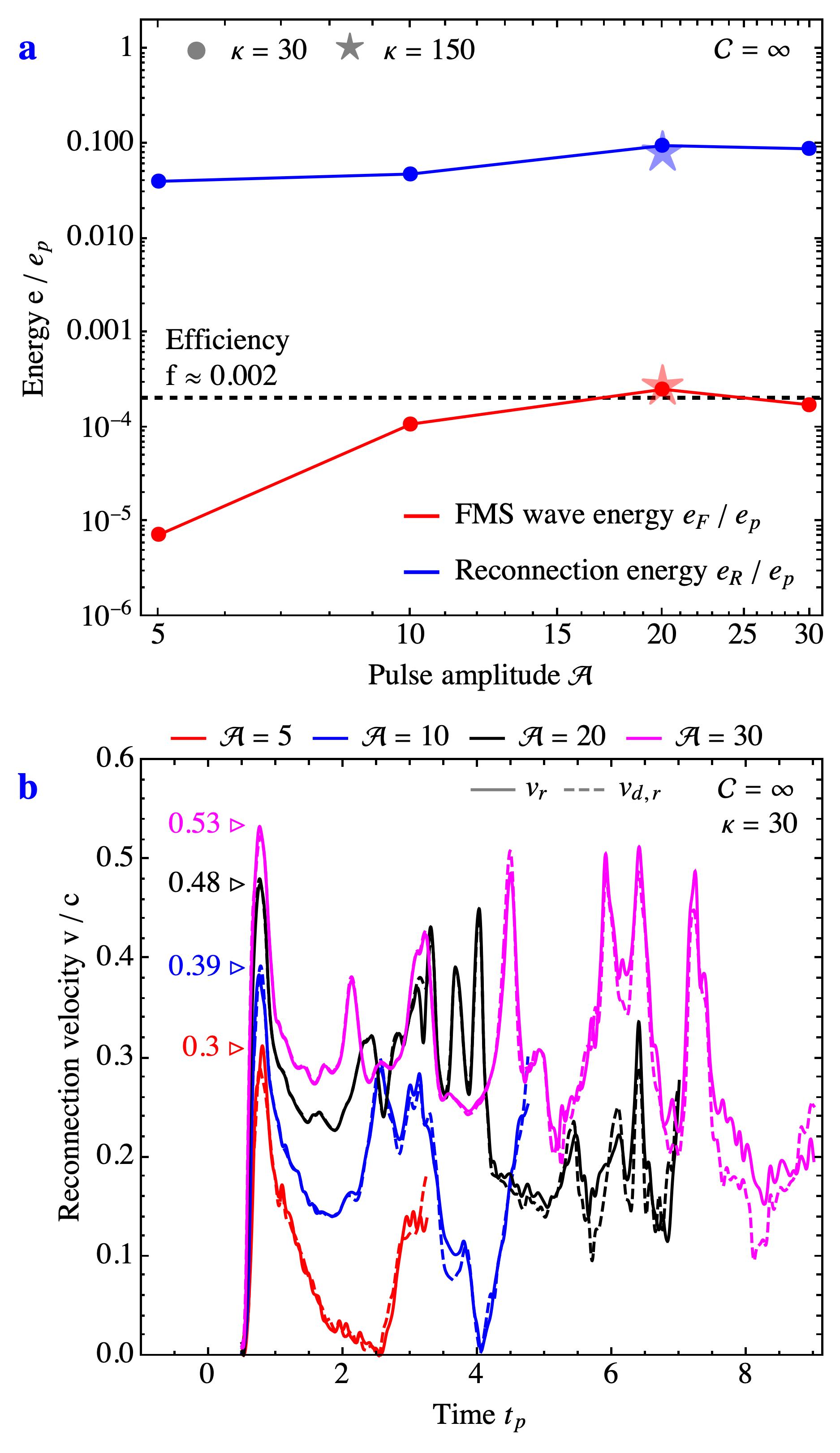}
  \vspace{-22pt}
  \caption{Energetics and reconnection rate during the LFP-current sheet interaction. We display the measured reconnected energy and the energy stored in the emerging high-frequency FMS waves ({a}). The reconnection velocity measured for different pulse amplitudes is shown in panel ({b}).}
\label{fig:energetics}
\end{figure}

To illustrate the plasma dynamics during the LFP-current sheet interaction, we track their evolution for $\sigma_{\rm u}=100$, $\kappa \approx 150$ and $\mathcal{A}=20$ to display its main stages in Figure~\ref{fig:largesystems}. One quantity that is used for the analysis shown in subsequent sections is the electric field of FMS waves induced by plasmoid mergers. Such waves correspond to the extraordinary (X) mode with the electric field perpendicular to the wavevector and background magnetic field, which in this setup corresponds to the $\tilde{E}_z$ component. The high-frequency plasma waves are imprinted on top of the LFP. To filter out the background fields, we subtract local averages calculated by smoothing the fields with a Gaussian kernel. Specifically, we use a kernel with a standard deviation of one skin depth $d_{\rm u}$ in the $y$-direction. We limit the window size to a narrow width along the $x$-direction. Starting from the initial configuration (Figure~\ref{fig:largesystems}{a}), the stable current sheet is compressed and its initial width of $a=20d_{\rm u}$ is reduced to Larmor radius scale (see Appendix B), triggering active reconnection. In Figure~\ref{fig:largesystems}{b}, small plasmoids appear and merge to form hierarchically larger ones while emitting high-frequency FMS waves during mergers. Plasmoid coalescence on top of the background pulse induces a strong wave signal. By this time, in Figure~\ref{fig:largesystems}{c}, a large fraction of the initial magnetic islands have merged. The large plasmoids accumulate a significant amount of particles, leading to a density decrease in the downstream of the LFP. In the wake of the main pulse, the plasma skin depth increases and one finds hotter plasma leaving the interaction region. With the large magnetic islands decoupling from the LFP, secondary current sheets form that are now directed along the pulse propagation direction (see vertical structures, Figure~\ref{fig:largesystems}{d}). The collisions of secondary structures induce more FMS waves that are then trailing behind the main signal. 

Figure~\ref{fig:energetics} is dedicated to the energetics of the compressed reconnection. For the extended box simulation ($\kappa\approx 150$) presented in this section, the LFP energy reconnected during the interaction is $e_{\rm r}/e_{\rm p}\approx 0.08$ (see star markers in Figure~\ref{fig:energetics}{a}). A fraction of this reconnected energy is transformed into high-frequency FMS waves, and we denote the conversion efficiency as $f=e_{\rm F}/e_{\rm r}\approx 2.5\times 10^{-3}$. Overall, the ratio between FMS wave energy and initially available LFP energy is $e_{\rm F}/e_{\rm p}\approx 2\times 10^{-4}$.

\subsubsection{LFP amplitude dependence}
\label{sec:amplitudedependence}

In this section, we verify the amplitude dependence of the energy conversion mechanism and FMS wave signal given by Equation~(\ref{eq:outgoinguncooled}). For this parameter scan, we choose $\sigma_{\rm u}=100$ and $\kappa\approx 30$, which we found to be sufficient to obtain similar results to the extended domain ($\kappa \approx 150$, see Section~\ref{sec:largesystems}) for reconnection rates and spectral signatures while minimizing computational cost. We consider the amplitudes $\mathcal{A}\in\left[5,10,20,30\right]$. Consistent with the scaling that we derived in Equation~(\ref{eq:recenergy}), the reconnected energy is at a level of $e_{\rm r}/e_{\rm p}\approx 0.1$ for all amplitudes (Figure~\ref{fig:energetics}{a}). Furthermore, we find that the efficiency of energy conversion into high-frequency FMS waves is $f\sim 0.002$, which effectively translates to $e_{\rm F}/e_{\rm p}\approx 2\times 10^{-4}$ \citep[see][]{Philippov2019}. Clearly, the low-amplitude ($\mathcal{A}=5$) data point is an outlier. In this case, reconnection is triggered by compression; however, most plasmoids merge after the current sheet has decoupled from the LFP. In other words, for low amplitudes the interaction time $t_{\rm i}$ is too short to capture FMS wave injection during compression.

To estimate the reconnection rate in the compressed layer one requires to compute the EM field components in the pulse frame. The respective Lorentz transformation relies on a reference velocity $v_{\rm p}$ of the background plasma. It can be derived by assuming that the total plasma velocity, $v$, in the laboratory frame consists of the sum of pulse velocity, $v_{\rm p}$, and reconnection velocity, $v_{\rm r}$. For relativistic plasma, this sum is
\begin{align}
    v=\frac{v_{\rm p}+v_{\rm r}}{1+v_{\rm p}v_{\rm r}/c^2}.
\end{align}
By measuring the total $x$-velocities $v^x_R$ and $v^x_L$ at both sides of the current sheet (left and right), as well as assuming that $v_{R,\rm r}=-v_{L,\rm r}$ and $v_{R,\rm p}=v_{L,\rm p}$, one finds a system of two equations. Such a system can then be solved to obtain the pulse and reconnection velocities:
\begin{align}
    \begin{split}
        v_{\rm p}&=\frac{c^2+v^x_L v^x_R-\sqrt{\left(c-v^x_R\right)\left(c+v^x_R\right)\left(c-v^x_L\right)\left(c+v^x_L\right)}}{v^x_L+v^x_R}
    \end{split}\\
    \begin{split}
        v_{\rm r}&=\frac{c^2-v^x_L v^x_R-\sqrt{\left(c-v^x_R\right)\left(c+v^x_R\right)\left(c-v^x_L\right)\left(c+v^x_L\right)}}{v^x_L-v^x_R}
    \end{split}\label{eq:reclab}
\end{align}
Alternatively, the velocity of the inflow into the current sheet in units of the Alfvén speed, $v_{d,\rm r}/c$, or `reconnection rate', can be calculated as
\begin{align}
    \frac{v_{d,\rm r}}{c}\approx\frac{E'_z}{B'_y}=\frac{E_z+v_{\rm p} B_y/c}{B_y+v_{\rm p} E_z/c}.
    \label{eq:recflds}
\end{align}
In practice, we average all fields and velocities along the $y$-direction prior to the velocity evaluation at a distance of $10\times d_{\rm u}$ from the location of the field reversal in $B_y$. In Figure~\ref{fig:energetics}{b}, we compare reconnection rates as derived with Eqs.~(\ref{eq:reclab}) and~(\ref{eq:recflds}) for different amplitudes. The data for all amplitudes are similar during active reconnection until the LFP and current sheet decouple. All cases display a rapid rise of the reconnection velocity up to a transient maximum value. This maximum value grows with the pulse amplitude. Subsequently, the reconnection velocity stabilizes at a lower value, $v_r\approx 0.2c$, until either the current sheet decouples from the LFP, or most plasmoids have merged.  

As we showed in the previous section, high-frequency FMS waves are induced by the subsequent coalescence of plasmoids of different sizes, and their spectral signature is a key observable. In the limit of slow synchrotron cooling, we found a $\hat{B}^{1/2}$ dependence of the peak frequency of outgoing FMS waves (Equation~\ref{eq:outgoinguncooled}). Figure~\ref{fig:spectra}{a} compares the energy spectra binned along the propagation direction ($k_x$) and reproduces the expected scaling (inset plot). By directly fitting Equation~(\ref{eq:outgoinguncooled}) to the data points, we can further estimate $\xi\zeta\sim 90$ for the parameters used to estimate the compressed current layer width and peak radiation frequency (Equation~\ref{eq:outfrequency}), showing good agreement with the assumptions of \citet{Lyubarsky2020}. 

We sample the outgoing FMS wave data for all simulations for each available slice in the $y$-direction to produce an individual dynamic spectrum. Such a spectrum resolves the energy bins of a 1D discrete Fourier transform over time (normalized by the pulse light-crossing time $t_{\rm p}=\Delta/c$) by shifting a sampling window (width of $10\;d_{\rm u}$) along the $x$-direction, and by assuming that the emerging fast waves travel with the speed of light along the same direction. The spectra assembled in Figure~\ref{fig:dynamicspectra} reveal the wave injections by showing coherent stripes of large intensity. As we illustrate in the left column of Figure~\ref{fig:largesystems}, plasmoid mergers during the induced reconnection event follow a hierarchy; the smallest plasmoids merge first, and consecutively feed coalescence of larger plasmoids. As the size of the merging plasmoids increases, the frequency shifts to lower values. Such frequency drifts (i.e., energy in high frequencies that fades into lower bands at later times) appear in some of the dynamic spectra in Figure~\ref{fig:dynamicspectra}. Mergers of the largest plasmoids in the wake of the LFP - now individually occurring rather than simultaneous events - are imprinted in the spectra by short-duration spikes at later times. Overall, the spectral signature of the low-amplitude compression ($\mathcal{A}=5$) is much weaker than in the other cases. Thus, our simulations require amplitudes $\mathcal{A}\gg 5$ to fully capture reconnection during strong compression (a central premise of this model). Finding very similar dynamic spectra for different box sizes ($\kappa\approx 30$ versus $\kappa\approx 150$) shows that the relevant dynamics and spectral features are, indeed, captured in the smaller domain.

\subsubsection{Effects of synchrotron cooling}
\label{sec:cooling}

\begin{figure}
  \centering
  \includegraphics[width=0.47\textwidth]{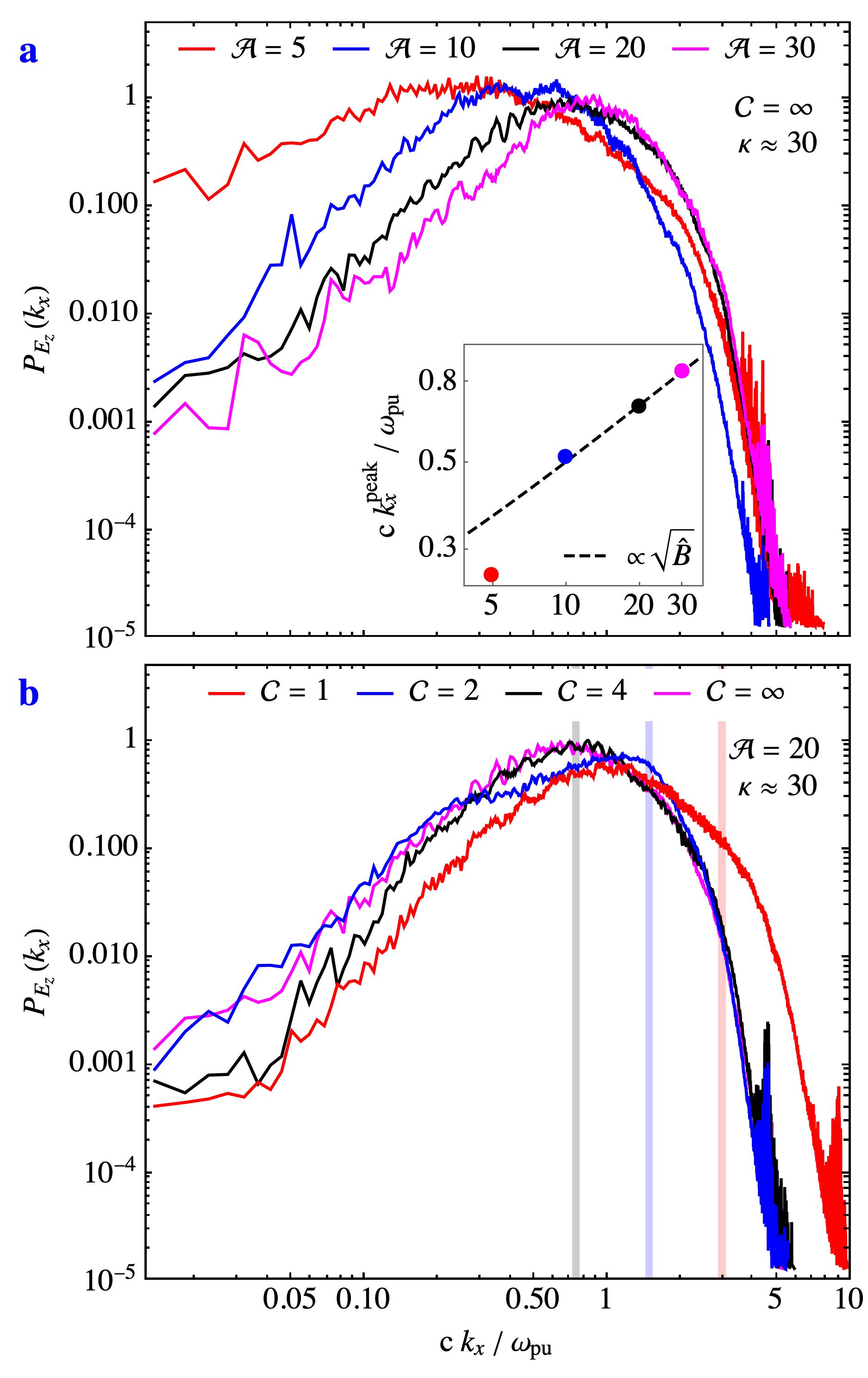}
  \vspace{-18pt}
  \caption{High-frequency FMS wave spectra with powers binned along the propagation direction ($k_x$) for different amplitudes $\mathcal{A}$ (panel {a}, Section~\ref{sec:amplitudedependence}), and different cooling parameters $\mathcal{C}$ (panel {b}, Section~\ref{sec:cooling}). The inset in ({a}) shows the peak energy as a function of wave amplitude. The vertical lines in ({b}) indicate an inversely proportional scaling of the frequency with $\mathcal{C}$ (see Eqs.~\ref{eq:outfrequency} and~\ref{eq:coolingtime}).}
\label{fig:spectra}
\end{figure}

\begin{figure*}
  \centering
  \includegraphics[width=1.0\textwidth]{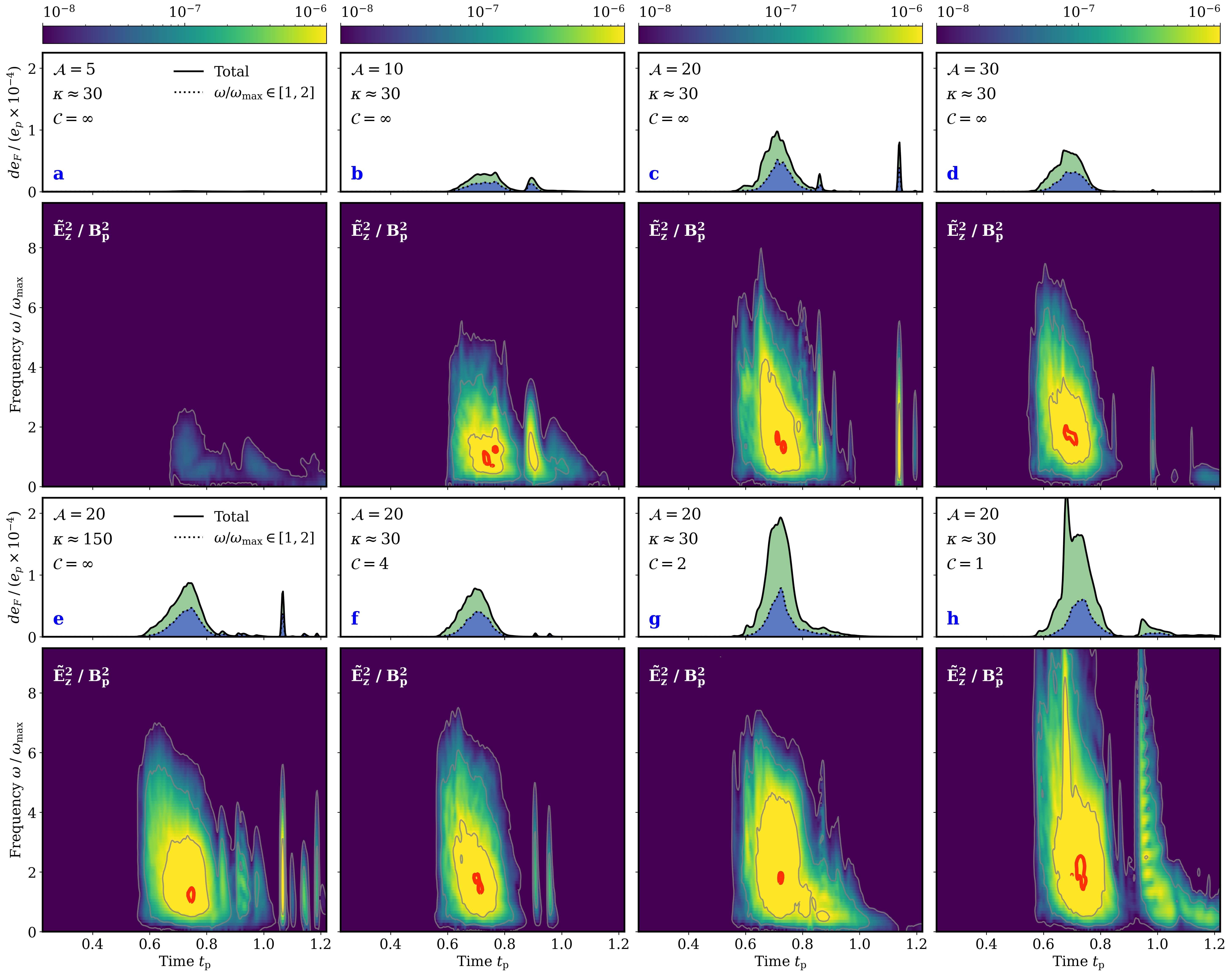}
  \vspace{-18pt}
  \caption{Dynamic spectra and power for different pulse amplitudes (top two rows; Section~\ref{sec:amplitudedependence}), $L_y$ extensions (bottom leftmost panels; Section~\ref{sec:largesystems}), and cooling parameters (bottom right panels; Section~\ref{sec:cooling}). The parameters of each model are given in the respective power plots on top of the dynamic spectra. Gray contours stacked in the spectra indicate differences by a factor of 10; red contours denote $90\%$ of the maximum energy. The frequency is normalized to the peak frequency $\omega_{\rm max}$ of the reference case with $\mathcal{A}=20$, $\kappa\approx 30$, and $\mathcal{C}=\infty$. 1D outlines on top of the spectra denote the corresponding power of the fast-wave signal (black solid line for total energy, black dotted line for the $\omega/\omega_{\rm max}\in [1,2]$ band).}
\label{fig:dynamicspectra}
\end{figure*}

In this section we test the influence of synchrotron cooling on the characteristic frequencies of FMS waves and interaction dynamics shown in Equation~(\ref{eq:outfrequency}). The strength of synchrotron cooling is parametrized by the ratio between the cooling time $t_{\rm c}$, and the particle acceleration time $t_{\rm a}$ at X-points \citep[see][]{Beloborodov2017a}:
\begin{align}
    \frac{t_{\rm c}}{t_{\rm a}}\approx\frac{F_{\rm a}}{F_{\rm c}}=\frac{8\pi e E'_{\rm rec}}{\sigma_{\rm T} \gamma'^2_{\rm rec} B'^2}=\frac{\gamma'^2_{\rm rad}}{\gamma'^2_{\rm rec}}\approx\frac{\gamma'^2_{\rm rad}}{\sigma'^2}\equiv\mathcal{C}^2.
    \label{eq:coolingtime}
\end{align}
Here, we employ the Thomson cross section $\sigma_{\rm T}$, and the typical particle Lorentz factor reached at X-points, $\gamma'_{\rm rec}$. Furthermore, $\gamma'_{\rm rad}$ is the Lorentz factor for which the radiation drag force is comparable to the accelerating force in the plasma rest frame, where we assume $E'_{\rm rec}=0.1 B'$ for the accelerating electric field \citep[see Section 2.2 in][]{Hakobyan2019}. For this parameter scan, we focus on the case of a pulse amplitude $\mathcal{A}=20$ in a $\kappa \approx 30$ domain with $\sigma_{\rm u}=100$. In the pulse rest frame, the magnetization is 
\begin{align}
    \sigma'=\frac{\hat{B}_{\rm p}}{\hat{\gamma}_{\rm p}}\gamma_{\rm u}\sigma_{\rm u}=\left(2\hat{B}_{\rm p}-1\right)^{1/2}\sigma_{\rm u}.
\end{align}
The upstream equivalent of $\gamma'_{\rm rad}$ is
\begin{align}
    \gamma_{\rm rad,u}=\left(\frac{\gamma'_{\rm rad}}{\sigma'}\right)\left(2\hat{B}_{\rm p}-1\right)^{3/4}\sigma_{\rm u}.
\end{align}
To explore the transition from weak to intermediate cooling, we explore the parameter range $\mathcal{C}\in\left[4.0,2.0,1.0\right]$. This corresponds to the fiducial values $\gamma_{\rm rad,u}\in\left[6240, 3120, 1560\right]$. According to Equation~(\ref{eq:outfrequency}), the main effect of synchrotron cooling is to shift the FMS wave spectra toward higher frequencies, with $\nu\propto 1/\mathcal{C}$. This is because strong cooling leads to the loss of pressure support of plasmoids, reducing their sizes and inducing higher frequency radiation during mergers \citep[see Appendix~\ref{sec:appendixC} and][]{Hakobyan2019}. Hence, the stronger the cooling, the harder it is to resolve the FMS waves on a finite numerical grid. We face this challenge by two adjustments to the setup described in Section~\ref{sec:SimulationSetup}. (i) We balance the particle load in strongly compressed plasmoids by a PIC particle merging algorithm (see Appendix~\ref{sec:appendixB}); and (ii) we choose to increase the resolution for the $\mathcal{C}=1$ simulation. This corresponds to resolving the skin depth by 100 cells while keeping the dimensionless scale parameters $\Delta/d_{\rm u}$, $a/d_{\rm u}$, and $\kappa$ the same. We adopt a lower number of particles per cell, $n_{\rm ppc, u}=1$, to decrease the computational cost.

Figure~\ref{fig:spectra}{b} shows the frequency distribution of the outgoing FMS waves for different cooling strengths. Compared to the limit of $\mathcal{C}\rightarrow\infty$ (see Figure~\ref{fig:spectra}{a}), synchrotron cooling becomes important for $\mathcal{C}\lesssim 4$, as to say there is a measurable shift in the signal's peak frequency in the intermediate cooling regime. The shift of the high-frequency component of the FMS waves approximately follows the expectation of inverse proportionality to the cooling strength $\mathcal{C}$, as one can see by comparing the peak of the spectra to the vertical lines in Figure~\ref{fig:spectra}{b} (plotted in intervals of factor two). Although the strongest cooling ($\mathcal{C}=1$, red curve) has a frequency peak that is comparable to the intermediate cooling strength ($\mathcal{C}=2$, blue curve), it shows a significant broadening to higher frequencies. The reconnection dynamics and overall dissipation is unaffected by cooling.

The dynamic spectra for the simulations with cooling are presented in the bottom panels of Figure~\ref{fig:dynamicspectra}. The noncooled reference case (panel e, $\kappa\approx 150$) is followed by the results with increasing cooling strength in panels (f)-(h). The spectral features identified in the no cooling limit (Section~\ref{sec:amplitudedependence}) persist: a high-frequency burst is followed by a downward frequency drift, following the hierarchically increasing plasmoid sizes.\footnote{Furthermore, Equation~(\ref{eq:outfrequencylab}) shows that for constant $\Gamma_{\rm p}$, the frequency of injected FMS waves decreases with increasing distance from $R_{\rm LC}$. For a magnetic field decaying as $\sim 1/r$ one finds $\nu\sim r^{-3/2}$, leading to an additional mechanism for the downward frequency drift.} Plasmoid mergers in the wake of the LFP induce secondary current sheets as well as short and bright high-frequency bursts of $\tilde{E}_z$, clearly visible at $t_{\rm p}\approx 1$ in panel (h). An overall shift to higher frequencies is the prominent characteristic of these cooled setups (illustrated by the larger amount of energy in high frequencies in panels (g)/(h) of Figure~\ref{fig:dynamicspectra}).

The results presented in this section illustrate the impact of synchrotron cooling during reconnection on observable FMS waves, namely, a natural shift toward higher frequencies. A realistically strong pulse in the outer magnetar magnetosphere will be strongly affected by synchrotron cooling, as we discuss below. 

\section{Discussion}
\label{sec:dicsussion}

The FRB model described in this work targets bursts with $\tau \sim t_p \sim 1\text{ms}$ duration, a total energy release in the radio band of $e_{\rm b}\sim 10^{39}\text{erg}$, and a luminosity of $L_{\rm b}\sim e_{\rm b}/\tau\approx 10^{42}\text{erg}\;\text{s}^{-1}$. Typical LFP luminosities associated to such FRBs are $L_{\rm p}\sim 10^{47}\text{erg}\;\text{s}^{-1}$. Our simulations show a conversion efficiency $f\sim 2\times 10^{-3}$ between reconnected magnetic field energy and high-frequency fast waves (Figure~\ref{fig:energetics}). The magnetic field at the light cylinder is approximated as dipolar, which yields
\begin{align}
    B_{\rm LC} = \frac{\mu}{R_{\rm LC}^3}=9.2\times 10^3\left(\frac{B_*}{10^{15}\text{G}}\right)\left(\frac{1\text{s}}{P}\right)^3\text{G}.
\end{align}
For a nearly spherical pulse, the pulse amplitude is
\begin{align}
\begin{split}
        B_{\rm p}&=\frac{1}{R_{\rm LC}}\left(\frac{L_{\rm p}}{c}\right)^{1/2}\\
        &=3.8\times 10^8\left(\frac{L_{\rm p}}{10^{47} \text{erg}\;\text{s}^{-1}}\right)^{1/2}\left(\frac{1\text{s}}{P}\right)\text{G}.
        \label{eq:pulseamp}
\end{split}
\end{align}
The relative amplitude of the pulse at the light cylinder, $b_{\rm LC}=B_{\rm p}/B_{\rm LC}$, is
\begin{align}
\begin{split}
        b_{\rm LC}\approx 4.1\times 10^4\left(\frac{L_{\rm p}}{10^{47} \text{erg}\;\text{s}^{-1}}\right)^{1/2}\left(\frac{10^{15}\text{G}}{B_*}\right)\left(\frac{P}{1\text{s}}\right)^{2}.
        \label{eq:amplitudeluminosity}
\end{split}
\end{align}
To obtain GHz frequencies for FMS waves injected close to the light cylinder, one finds with Equation~(\ref{eq:outfrequencylab})
\begin{align}
\begin{split}
        b_{\rm LC}&\approx4.2\times 10^{4}\left(\frac{\beta_{\rm rec}}{0.1}\right)^{2/5}\left(\frac{\xi\zeta}{100}\right)^{4/5}\\
        &\qquad\left(\frac{\nu}{1\text{GHz}}\right)^{4/5}\left(\frac{10^{15}\text{G}}{B_*}\right)^{6/5}\left(\frac{P}{1\text{s}}\right)^{18/5}.
        \label{eq:amplitudefrequency}
\end{split}
\end{align}
Both constraints in Eqs.~(\ref{eq:amplitudeluminosity}) and (\ref{eq:amplitudefrequency}) are required to explain FRBs induced by the mechanism described in this paper. The strength of synchrotron cooling and relativistic bulk motion of particles in the interaction region control the characteristic frequency of the wave emission \citep[][]{Lyubarsky2020}.
In the case of the finite system, combining Eqs.~(\ref{eq:erfinite}) and~(\ref{eq:pulseamp}), the total burst luminosity can be written as
\begin{align}
\begin{split}
    & L_{\rm b}\approx \frac{f}{\tau}e_{\rm r}=\frac{f}{\tau}\left(\frac{L_{\rm p}}{c}\right)^{1/2}B_{\rm LC}R_{\rm LC}^2\\
    &=10^{42}\left(\frac{L_{\rm p}}{10^{47} \text{erg}\;\text{s}^{-1}}\right)^{1/2}\left(\frac{B_*}{10^{15}\text{G}}\right)\left(\frac{1\text{s}}{P}\right)\left(\frac{1\text{ms}}{\tau}\right)\text{erg}\;\text{s}^{-1}.
    \label{eq:totalenergyrelease}
\end{split}
\end{align}
$L_{\rm b}$ scales with the inverse of the rotational period and FRB duration. Bright millisecond FRBs produced by this mechanism may be related to faster rotating magnetars. Combining Eqs.~(\ref{eq:outfrequencylab}), (\ref{eq:amplitudefrequency}), and (\ref{eq:totalenergyrelease}) yields
\begin{align}
\begin{split}
    % e_{\rm b}&\approx 2.1\times 10^{40}\left(\frac{\beta_{\rm rec}}{0.1}\right)^{9/5}\left(\frac{\tau}{1\text{ms}}\right)\left(\frac{\xi\zeta}{100}\right)^{8/5}\\
    %   &\qquad\left(\frac{\nu}{1\text{GHz}}\right)^{8/5}\left(\frac{10^{15}\text{G}}{B_*}\right)^{2/5}\left(\frac{P}{1\text{s}}\right)^{16/5}\text{erg}.
    %     \nu&\approx0.6\times\left(\frac{e_{\rm b}}{10^{40}\text{erg}}\right)^{5/8}\left(\frac{B_*}{10^{15}\text{G}}\right)^{1/4}\left(\frac{1\text{s}}{P}\right)^2\\
    % &\qquad \left(\frac{0.1}{\beta_{\rm rec}}\right)^{9/8}\left(\frac{1\text{ms}}{\tau}\right)^{5/8}\left(\frac{100}{\xi\zeta}\right)\text{GHz.}
    \nu&\approx 1\times\left(\frac{L_{\rm b}}{10^{42}\text{erg}\;\text{s}^{-1}}\right)^{5/4}\left(\frac{10^{15}\text{G}}{B_*}\right)\left(\frac{1\text{s}}{P}\right)^{3/4}\\
    &\qquad \left(\frac{0.1}{\beta_{\rm rec}}\right)^{1/2}\left(\frac{100}{\xi\zeta}\right)\left(\frac{\tau}{1\text{ms}}\right)^{5/4}\text{GHz.}
       \label{eq:frequencyluminosity}
\end{split}
\end{align}
Coherent emission emerging during the LFP-current sheet interaction in the magnetar wind falls into the GHz range for intermediate to high burst luminosities, $L_{\rm b}\gtrsim 10^{42} {\rm erg/s}$, for magnetars with periods as observed in our Galaxy, or younger magnetars with faster rotation.

It is important to note that the radiation of FMS waves ceases if the opposite magnetic flux provided by the magnetar wind is insufficient and, hence, reconnection stops.\footnote{In addition to a vanishing upstream flux, the FMS wave injection can cease when the expansion of the magnetar wind outperforms the plasmoid coalescence. Such a ``freezing'' of plasmoid mergers can happen beyond $r\sim R_{\rm LC}/\beta_{\rm rec}$.} With the heuristic arguments provided in Section~\ref{sec:PlasmaScaling}, this can happen if reconnection time, $t_{\rm r}\sim (1/\beta_{\rm rec})\times R_{\rm LC}/c$,  is shorter than the interaction time given by Equation~(\ref{eq:interaction}), $t_i\sim(b_{\rm LC}/2) \times (\Delta/c)$. The ratio of these timescales can be calculated as $\mathcal{F}=t_r/t_i\sim 2/\beta_{\rm rec}\times \phi_{\rm u}/\phi_{\rm p}$, where we used $\phi_{\rm p}/\phi_{\rm u}\sim b_{\rm LC}\times (\Delta/R_{\rm LC})$. In Appendix~\ref{sec:appendixD}, we present an adaptation of the simulations analyzed in Section~\ref{sec:amplitudedependence} that mimics a finite upstream flux reservoir with $\mathcal{F}\approx 0.5$. We find that the proposed emission mechanism continues to operate and produce radiation at similar frequencies, while intensity and duration of the FMS wave packet is moderately reduced compared to the case of infinite system (see Figure~\ref{fig:finiteflux}). In reality, we expect a ratio of
\begin{align}
    \mathcal{F}\approx 0.1\times\left(\frac{4\times 10^4}{b_{\rm LC}}\right)\left(\frac{1\text{ms}}{\tau}\right)\left(\frac{0.1}{\beta_{\rm rec}}\right)\left(\frac{P}{1\text{s}}\right).
    \label{eq:fluxlimit}
\end{align}
The limited upstream flux may lead to a shorter radio signal, $\tau \lesssim \Delta/c \sim 1\text{ms}$. However, the FRB duration is likely to be less tightly constrained for several reasons. For example, the effective gravity in the accelerating wind may lead to mixing of the current sheet into the body of LFP due to the Kruskal-Schwarzschild instability \citep{Lyubarsky2010,Gill2017}. Also, the LFP-current sheet interaction is not planar on a global scale. Rather, the LFP reaches the reconnection layer at different times. Consequently, the distribution of FMS waves relative to the pulse width in both space and time are likely to counteract the finite flux limit by accumulating subpulses to form the resulting FRB.

Different scenarios of triggering reconnection in the outer magnetar magnetosphere can still produce GHz bursts but are not bound to have the relation between burst luminosity and peak frequency in Equation~(\ref{eq:frequencyluminosity}). For example, \citet{Yuan2020} proposed that escaping large-amplitude Alfvén waves trigger reconnection in their tail within $R_{\rm LC}$, where GHz emission produced by merging plasmoids is possible without field compression. It is unclear how the opening-up of the magnetosphere affects the propagation and the growth of the relative amplitude of these FMS waves. Processes like the damping of high-amplitude FMS waves discussed by \citet{Beloborodov2021a}, driven by nonideal plasma effects, can further affect signals propagating within $R_{\rm LC}$. Observations constraining a possible FRB frequency-luminosity relation and identifying populations that scale according to Equation~(\ref{eq:frequencyluminosity}) will ultimately probe the presented model.

Synchrotron cooling can be significant for a strong magnetic pulse close to the magnetar light cylinder. The strength of cooling is parametrized by the ratio $\gamma'_{\rm rad}/\sigma'$, 
\begin{align}
    \gamma'_{\rm rad}=\left(\frac{3}{2}\beta'_{\rm rec}\frac{B_{\rm cl}}{B'_0}\right)^{1/2},
\end{align}
and $B_{\rm cl}=m_e^2 c^4/e^3$; $B'_0=b_{\rm LC}B_{\rm LC}/\Gamma_{\rm p}$. Using the estimates of \citet[][Eqs.~17;~22]{Lyubarsky2020} for the density in the pulse,
\begin{align}
\begin{split}
    \mathcal{C}=\frac{\gamma'_{\rm rad}}{\sigma'}>&\; 4.6\times 10^{-6}\left(\frac{10^{15}\text{G}}{B_*}\right)^{1/18}\left(\frac{\beta_{\rm rec}}{0.1}\right)^{1/2}\\
&\left(\frac{10^{47} \text{erg}\;\text{s}^{-1}}{L_{\rm p}}\right)^{1/2}\left(\frac{\mathcal{M}}{10^3}\right)^{2/3}\left(\frac{P}{1\text{s}}\right)^{1/2}.
\label{eq:realsync}
\end{split}
\end{align}
This estimate is a lower limit on $\mathcal{C}$ as the pulse magnetization depends on the plasma loading to which pair production, by photons produced by accelerating particles in the current sheet, can contribute significantly. For any $\mathcal{C}<1$, the scaling of Equation~(\ref{eq:coolingtime}) holds.

During the interaction, a nonnegligible fraction of the incident pulse energy is dissipated in a compact region. The magnetic compactness in the compression zone \citep[][]{Beloborodov2021} can be approximated by using $R_{\rm LC}$ as the characteristic length scale of the emitting region ($S$, transverse to the propagation direction in this scenario), and the magnetic energy $U'_B=B'^2/(8\pi)$:
\begin{align}
\begin{split}
    l'_B&= \frac{\sigma_{\rm T} U'_B S}{m_e c^2}=\Gamma_{\rm p}^{-2}\frac{\sigma_{\rm T} B_{\rm p}^2 S}{8\pi m_e c^2}\\
    &\approx 2.1\times 10^3\left(\frac{b_{\rm LC}}{4\times 10^4}\right)\left(\frac{B_*}{10^{15}\text{G}}\right)^2\left(\frac{1\text{s}}{P}\right)^5.
\end{split}
\end{align}
The order of magnitude of $l'_B$ is comparable to flares expected in magnetically dominated accretion disk coronae \citep{Beloborodov2017a}. It suggests that radiative processes become (dynamically) relevant, and are likely to produce hard X-ray spectra. Enhanced pair production is likely to occur in this regime and should be evaluated in a separate study. While such pair loading will decrease the magnetization $\sigma'$ and can affect the LFP propagation, the underlying mechanism of fast-wave injection is universal. The FMS frequency and pulse duration do not depend on plasma characteristics as long as $\sigma'>1$ and $\mathcal{C}<1$.

Observations of nonthermal X-ray radiation from the galactic magnetar \object{SGR J1935+2154} during \object{FRB 200428} find ratios between peak radio to peak X-ray luminosity of $\lesssim 10^{-3}$ and between radio to X-ray energy of $\lesssim 10^{-5}$ \citep{Mereghetti2020,Ridnaia2021}. The numerically constrained process efficiency $f\sim 0.002$ (Figure~\ref{fig:energetics}{a}) is a conservative upper limit for the expected ratio of radio emission to X-ray production. The comparison to the aforementioned observations has limited validity as \object{FRB 200428} is not in the range of suitable luminosities and frequencies of this model (Equation~\ref{eq:frequencyluminosity}). A detailed examination of the expected high-energy signals will be conducted in future work.

\object{FRB 20200120E} is another interesting recent observation, identifying a 100 nanosecond substructure in the burst signal \citep{Majid2021}. Such short-duration dynamics can be produced by the reconnection-mediated FRB model when a hierarchy of individual nanoshots from large plasmoid mergers form a packet of high-frequency fast waves with nonoverlapping components. Especially, mergers occurring during the decompression phase induce well-separated bursts (Figure~\ref{fig:dynamicspectra}).

\section{Conclusions}
\label{sec:conclusion}

In this work, we present the first comprehensive numerical simulations of a \emph{reconnection-mediated} model for FRB generation in the outer magnetar magnetosphere \citep{Lyubarsky2020,Lyubarsky2021}. Section~\ref{sec:PlasmaScaling} dissects the envisioned scenario into a series of subprocesses: LFP propagation, current sheet compression, reconnection, formation of plasmoids, and effects of strong synchrotron cooling. The underlying coherent emission mechanism is essentially the injection of fast waves by merging plasmoids \citep{Lyubarsky2019,Philippov2019}. As such, the large-scale 2D PIC simulations presented in Section~\ref{sec:Simulations} broadly confirm the scaling relations expected from theory. The reconnected energy is converted to energy transported by the outgoing high-frequency FMS waves with an efficiency factor $f\sim 0.002$. In the pulse rest frame, these waves appear as spherical fronts, emerging like fireworks from the coalescing plasmoids in the right panels of Figure~\ref{fig:largesystems}. The hierarchy of plasmoid sizes during reconnection of the compressed current sheet naturally induces a downward frequency drift of the FMS waves at later times of the interaction. Synchrotron cooling shifts the outgoing wave frequency to higher values, which confirms the theoretical picture by \citet{Lyubarsky2020,Lyubarsky2021}. We find that this scenario can explain the required properties of intermediate-to-high-luminosity FRBs (Section~\ref{sec:dicsussion}). 

The conducted PIC simulations are 2D. In future work, the process of wave injection during coalescence of 3D plasmoids needs to be validated in a separate survey of full 3D simulations. Dimensionality is also important for modeling the nonlinear interaction of waves in highly magnetized plasma, which may potentially affect the outgoing radiation signal \citep[e.g.,][]{Lyubarsky2020,Ripperda2021}. However, in this work we show that these questions can be addressed by probing aspects of the emission mechanism using local reconnection studies, instead of the full, numerically very demanding, LFP-current sheet interaction simulations. To conclude, we are confident that this work establishes the consistency of the reconnection-mediated FRB model by \citet{Lyubarsky2020}. 

\section*{Acknowledgments}

 We thank Bart Ripperda for testing our setup in their MHD code \textsc{BHAC}, and Miguel A. Aloy for their insight into the corresponding MHD discontinuities. We appreciate the help by Daniel Grošelj, who assisted in optimizing the dispersive properties of our field solver, and Joonas Nättilä, who shared their insight into stabilizing dispersion errors of highly nonlinear waves in PIC. We also thank Lorenzo Sironi and Chris Thompson for insightful discussions related to this model. We welcome the improvements to the discussion of our model suggested by a thorough referee. This research is supported in part by NASA grant 80NSSC18K1099 and NSF grant PHY-1804048. A.A.P. and J.F.M. acknowledge support from the National Science Foundation under grant No. AST-1909458. A.L. acknowledges support from the Israel Science Foundation grant 1114/17. This research is part of the \textit{Frontera} \citep{Stanzione2020} computing project at the Texas Advanced Computing Center (LRAC-AST21006). \textit{Frontera} is made possible by National Science Foundation award OAC-1818253. The presented numerical simulations were further enabled by the \textit{MareNostrum} supercomputer (Red Española de Supercomputación, AECT-2021-1-0006), and the \textit{Stellar} cluster (Princeton Research Computing). Research at the Flatiron Institute is supported by the Simons Foundation.

\bibliography{literature.bib} 

\appendix

\section{A magnetic low-frequency pulse in the highly magnetized limit}
\label{sec:appendixA}

\citet{Lyubarsky2020} reviews the propagation dynamics of an LFP as a nonlinear wave in a highly magnetized wind. We simplify such a nonlinear pulse to a Cartesian geometry with a constant background field $B_y=B_0$ along the $y$-direction. The continuity equation and Alfvén's theorem have congruent forms, namely

\begin{align}
    \partial_t N&+\partial_x\left(N v\right)=0,\label{eq:continuity}\\
    \partial_t B_y&+\partial_x\left(B_y v\right)=0.\label{eq:frozenin}
\end{align}

Here,  $v$ is the velocity along the $x$-direction and $N$ is the lab frame mass density. Direct comparison of Eqs.~(\ref{eq:continuity}) and~(\ref{eq:frozenin}) implies $B_y=\chi N$ with a suitably chosen constant $\chi$; the ratio $B_y/N$ is uniform along the pulse. Energy conservation in the cold plasma limit yields nonlinear pulses with 
\begin{align}
\frac{\text{d}\gamma}{\text{d} N}=\frac{v\gamma s_{\rm fast}/c}{N\left(1+v s_{\rm fast}/c^2\right)}\label{eq:gammadensity},
\end{align}
where
\begin{align}
    s_{\rm fast}/c=\left(\frac{\sigma}{1+\sigma}\right)^{1/2} .
\end{align}
For the case considered in this paper, namely, in the highly magnetized limit, $s_{\rm fast}/c\approx 1$. With a background fluid at rest, one finds 
\begin{align}
\frac{\text{d}\gamma}{\text{d} \hat{N}}=\frac{v\gamma }{\hat{N}\left(1+v/c\right)}\label{eq:gammadensityreduced}.
\end{align}
Here, we used $\hat{N}=N/N_0$, where $N_0$ is the background rest mass density. Equation~(\ref{eq:gammadensityreduced}) has an exact solution:
\begin{align}
    \gamma_{\rm p}=\hat{N}\left(2\hat{N}-1\right)^{-1/2}=\hat{B}\left(2\hat{B}-1\right)^{-1/2}
    \label{eq:gammapulseNRwind}
\end{align}
In the last equation, we substituted $\hat{B}=B/B_0$, where $B_0$ is the background magnetic field. Solving for the velocity of the pulse one is left with the drift velocity of the pulse $v_{\rm p}=(\hat{B}-1)/\hat{B}$.

\section{1D-PIC simulations of the LFP-current sheet interaction}
\label{sec:appendixA1}

\begin{figure}
  \centering
  \includegraphics[width=0.47\textwidth]{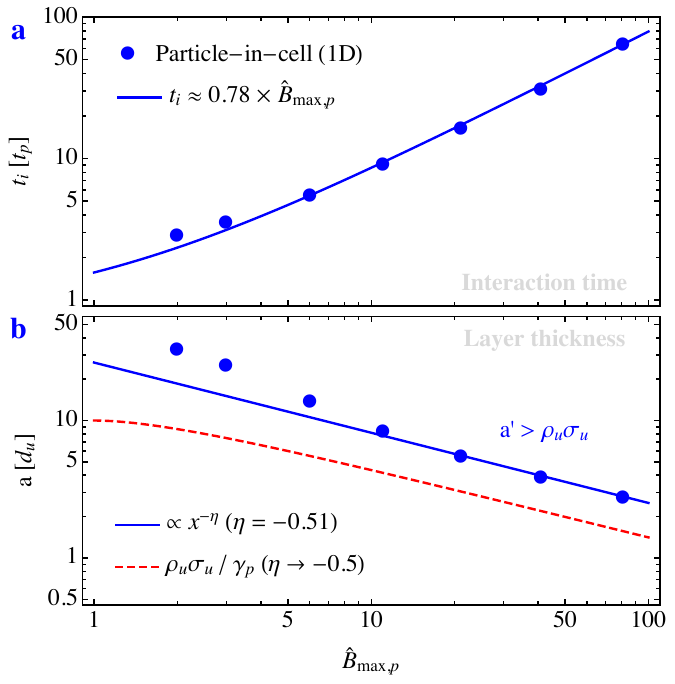}
  \vspace{-11pt}
  \caption{Time and length scales of the LFP-current sheet interaction in 1D-PIC simulations (using a setup as described in Section~\ref{sec:SimulationSetup}). We verify the scaling derived for the LFP-current sheet interaction time (panel {a}, see Equation~\ref{eq:interaction}) as well as the layer thickness (panel {b}, see Equation~\ref{eq:alphasecond}). We show the validity of the assumption $a'>\rho_{\rm u}\sigma_{\rm u}$ that implies the presence of sufficient charge carriers in the reconnection layer.}
\label{fig:1Dscaling}
\end{figure}

We use the numerical setup described in Section~\ref{sec:SimulationSetup} to conduct 1D-PIC simulations of the LFP-current sheet interaction. With reduced dimensionality, no reconnection physics can be captured. However, the 1D models provide insight into the interaction duration as well as the compression of the current sheet. We present the dependence of interaction time and compression of the reconnection layer width as a function of the pulse amplitude in Figure~\ref{fig:1Dscaling}. The collected results show the validity of Equation~(\ref{eq:interaction}) as well as of the assumption used for the charge-starvation analysis in Equation~(\ref{eq:alphasecond})

\section{Current-conserving PIC particle merging}
\label{sec:appendixB}

\begin{figure*}
  \centering
  \includegraphics[width=0.98\textwidth]{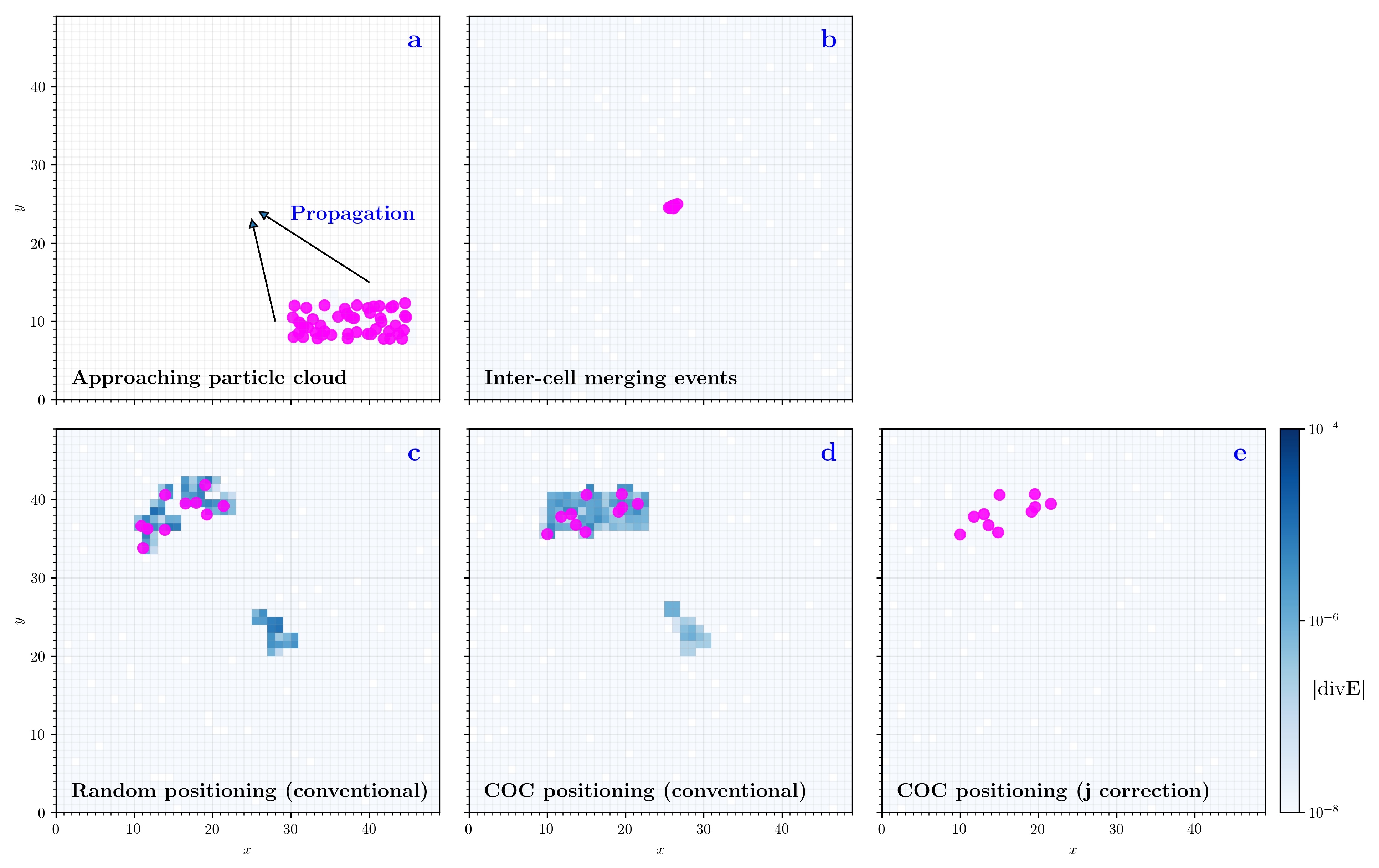}
  \vspace{-6pt}
  \caption{Illustration of different macroparticle merging algorithms at the example of an initial charge-neutral particle cloud with trajectories that intersect in one cell, triggering merger events. We display the merging particles (magenta colored dots) as well as the local charge density $|\text{div}\mathbf{E}|$ (blue color density) at three different times. After the particle cloud approached the common intersection cell ({a}) and particles are merged consecutively ({b}), different numerical techniques show different qualities of conserving charge neutrality (c)/(d)/(e). Merged superparticles can be randomly positioned ({c}), positioned at the center of mass (COC) of precursor particles ({d}), or injected at the COC with a correction of their displacement in the current ({e}).}
\label{fig:merging}
\end{figure*}

Modeling plasma processes with PIC codes relies on a sufficiently fine sampling of phase space, effectively achieved by simulating macroparticles with continuous positions and momenta. Merging macroparticles in regions of extremely high densities like the ones that easily develop in the setup described in this work is a useful (supplementary) tool to limit memory usage to feasible loads\footnote{An alternative to the resampling of particle loads on fixed slabs of the grid is to optimize and adapt the subdomains assigned to each processor of a parallel infrastructure, as has been done, e.g., by \citet[][]{Naettilae2019}.}. Conservative PIC particle merging algorithms are one way to down-sample the particle load and were initially reviewed by \citet[][]{Assous2003}. Weight optimization techniques for merged particles have since been developed to conserve essential moments of the distribution function \citep{Welch2007,Faghihi2020}. \citet{Vranic2015} and \citet{Luu2016} hold a credible benchmark for their particle merging algorithm based on the proximity of particles in 6D phase space to simultaneously conserve charge, momentum, and energy of particles. It has been further developed and implemented into various state-of-the-art PIC codes \citep{Derouillat2018,tristanv2}. However, \citet{Gonoskov2020} notes that despite extensive efforts, commonly even minor charge relocation that is not supplemented with corresponding currents show up as noise in the EM field evolution. 

We address this challenge of current conservation by providing suitable currents to compensate inaccuracies during particle merging by adding three adaptations to the method discussed by \citet[][]{Vranic2015}. First, we only merge particles located in the same numerical cell, causally connected by a suitable CFL condition. Second, we place the newly merged particles at the center of charge (COC) of their progenitors. In applications where all particles have the same charge-to-mass ratio, the COC coincides with the center of mass. The current induced by moving PIC particles is given by the sum of the products of their charge $q_i$, and velocity $\mathbf{v}_i$. Naively approximated, an instantaneous shift of charged particles to the position of their COC (during the infinitesimal time $\varepsilon$) has no contribution to the corresponding current
\begin{align}
    \mathbf{j}=\sum_i q_i \mathbf{v}_i = \sum_i  \frac{q_i}{\varepsilon}\left(\mathbf{x}_i-\mathbf{x}_{\rm COC}\right)=\mathbf{0},
\end{align}
and, thus, the COC initialization seems to be a favorable choice. However, each particle's contribution to the discretized currents depends on the overlap of the corresponding shape function with the cell edges and on the specific particle path $\mathbf{x}_i-\mathbf{x}_{\rm COC}$. Therefore, finally, we ensure current conservation by depositing the current induced by an instantaneous particle shift to the COC after the particle merging step \citep[applying the zigzag algorithm; see][]{Umeda2003}. 

Figure~\ref{fig:merging} illustrates the effect of the proposed adaptations on conventional charged particle merging algorithms \citep[e.g.,][]{Vranic2015}. We follow an initially charge-neutral particle cloud consisting of positive and negative charges moving on top of each other (Figure~\ref{fig:merging}{a}) with one particle species undergoing consecutive merger events at an intersection point (Figure~\ref{fig:merging}{b}). Without explicit current corrections, the merger events - while conserving the global particle weight and charge - induce local electric field divergences. Positioning the newly merged particles at the COC notably reduces the magnitude of leftover divergences ({d}). However, only adding current corrections for all particle shifts during the merger events reduces the electric field divergences by several orders of magnitudes ({e}). While maintaining the implementation of local particle coalescence as introduced and assessed by \citet{Vranic2015}, we are confident that our enhanced current conservation techniques can significantly reduce noise in the EM mesh quantities. 

Current-conserving PIC particle merging was employed to handle the large compression of the synchrotron cooled plasmoids modeled in Section~\ref{sec:cooling}. In our simulations, the down-sampling pipeline is only called whenever the number of particles in one cell is $n/n_u>2000$. Particle momenta are distributed across $n_\theta\times n_\phi = 18\times 32$ spherical directional bins and $n_e=16$ logarithmically scaled energy bins. All particles that share the same bin are considered to be close in phase space and undergo current-conservative merging when the weight $w$ of the particles is $w /w_0<32$. These parameters are chosen to merge particles only in the most extreme density accumulations of large magnetic islands while avoiding excessive imbalances of the particle weight distribution.

\section{The equilibrium of synchrotron cooled plasmoids}
\label{sec:appendixC}

\begin{figure*}
  \centering
  \includegraphics[width=0.98\textwidth]{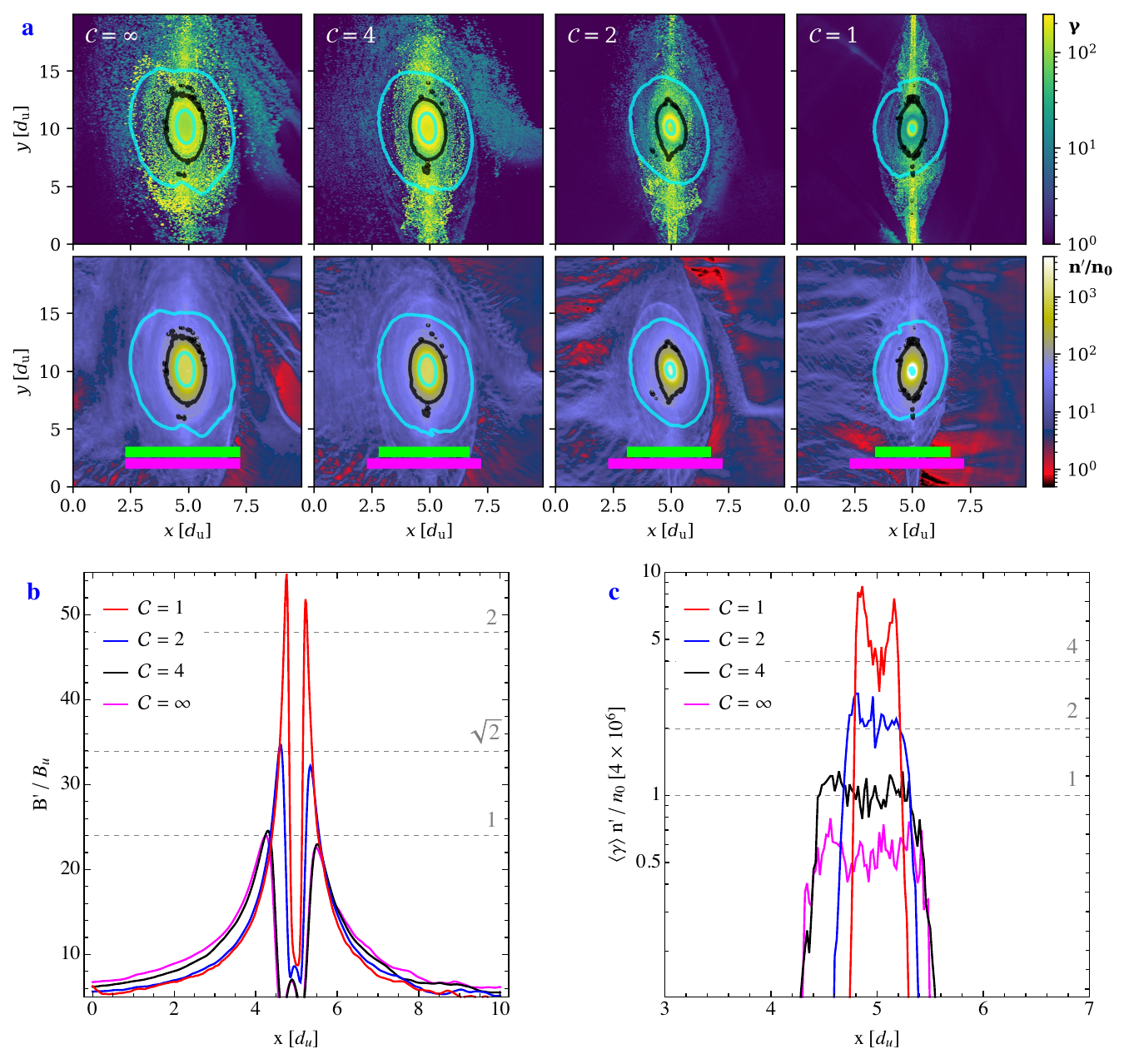}
  \vspace{-6pt}
  \caption{Selected representative plasmoids extracted from the simulation data at $t=4t_{\rm p}$ for different cooling strengths (as indicated). We analyze the particle Lorentz factor measured in the rest frame of the pulse plasma (panel {a}, top row) and the plasma density (panel {a}, bottom row). In panel ({a}), we indicate locations where $n'/n_0=10^2$ by a black contour in each panel, and locations with $B'/B_{\rm u}=10$ by cyan contours. The inset colored bars (panel {a}, bottom row) denote the size of the noncooled reference plasmoid (magenta), and the extent of each plasmoid (green), respectively. The size of characteristic plasmoids decreases with increasing strength of synchrotron cooling. Panels ({b}) and ({c}) show 1D outlines of selected quantities along the $y=10 d_{\rm u}$ coordinate (see panel {a}). First, we compare the magnetic field strength for different cooling strengths ({b}). The maximum magnetic field strength scales with approximately $\mathcal{C}^{-1/2}$. Finally, we repeat the comparison for the pressure term $\left\langle\gamma\right\rangle n'$ in ({c}) and find a scaling with approximately $\mathcal{C}^{-1}$.}
\label{fig:figureA2}
\end{figure*}

We examine the internal structure of plasmoids by assuming a balance of plasma pressure $P$, and magnetic forces at each radial distance from the structure's center \citep[see Appendix~A,][]{Hakobyan2021}. On fluid scales within the plasmoid shell, the magnetic field is assumed to be purely toroidal, as to say $\mathbf{B}'=B'^\phi(r)\boldsymbol{\hat{\phi}}$, where $(\boldsymbol{\hat{r}},\boldsymbol{\hat{\phi}},\boldsymbol{\hat{z}})$ is the orthogonal base in cylindrical coordinates. We, thus, obtain the following pressure equilibrium:
\begin{align}
    \frac{1}{c}\mathbf{j}'\times\mathbf{B}'=\nabla P'\qquad\qquad\Longleftrightarrow\qquad\qquad 
    % \frac{1}{8\pi}\partial_r\left[\left(r B_\phi\right)^2\right]=-r^2\partial_r P
    \partial_r\left[\frac{1}{2}\left(B'^\phi\right)^2+4\pi P'\right]=\frac{1}{r}\left(B'^\phi\right)^2.
    \label{eq:pressurebalance}
\end{align}
In analogy to \citet{goldston2020}, we integrate Equation~(\ref{eq:pressurebalance}) to find
\begin{align}
    4\pi P'(r)=4\pi P'_0-\frac{1}{2}\left(B'^\phi\right)^2-\int_0^r \frac{1}{r}\left(B'^\phi\right)^2 \text{d}r.
\end{align}
The magnetic field profile $B'^\phi(r)$ as well as the pressure $P'_{\rm out}$ at the edge of the current column determines the pressure $P'_0$ at the center of the plasmoid. Assuming a z-pinch profile with $B'^\phi(r)=2B'^\phi_{\rm out}r_{\rm out} r/(r^2+r_{\rm out}^2)$, we find
\begin{align}
    P'_0=P'_{\rm out}+\frac{3}{8\pi}\left(B'^\phi_{\rm out}\right)^2\qquad\Longrightarrow\qquad P'_0\approx \frac{3}{8\pi}\left(B'^\phi_{\rm out}\right)^2,
    \label{eq:zpinchpressure}
\end{align}
where we use $P'_0\gg P'_{\rm out}$ in the last approximation. Equation~(\ref{eq:zpinchpressure}) illustrates that the pressure at the origin of the plasmoid is expected to scale proportional to the magnetic pressure. 

Figure~\ref{fig:figureA2} assembles characteristic rest-frame properties for plasmoids found in the simulations with synchrotron cooling (Section~\ref{sec:cooling}) at $t=4t_{\rm p}$. At this time, plasmoids have completed two to three plasmoid mergers. Figure~\ref{fig:figureA2}{a} shows the decrease in plasmoid size for increased cooling strength (see green inset bars). The magnetic field contours (cyan), as well as the magnetic field density distribution in Figure~\ref{fig:figureA2}{b} reproduce an approximate z-pinch profile as used at the beginning of this section. With the combination of Figure~\ref{fig:figureA2}{b} and Figure~\ref{fig:figureA2}{c} we aim at probing the following pressure balance:
\begin{align}
    \left\langle\gamma'\right\rangle n'\sim\left(B'^\phi_{\rm out}\right)^2.\label{eq:coolingbalance}
\end{align}
The scaling of both sides of this equation with the cooling parameter $\mathcal{C}$ is consistent (see dashed gray lines in Figure~\ref{fig:figureA2}{b}/{c}). In other words, for plasmoids of a similar boundary magnetic field $B'^\phi_{\rm out}$, the product $\left\langle\gamma'\right\rangle n'$ is constant. For such plasmoids, increasing the strength of synchrotron cooling will decrease $\left\langle\gamma'\right\rangle$ and increase its maximum density. The rest-frame skin depth $d'$ scales as \citep[see][]{Hakobyan2019}
\begin{align}
    d'=\left\langle\gamma'\right\rangle^{1/2}\left(\frac{n'_u}{n'}\right)^{1/2} d'_{\rm u}\propto \mathcal{C}\times d'_{\rm u}.
\end{align}
According to Equation~(\ref{eq:coolingbalance}), stronger magnetic field compression further decreases the skin depth in the plasmoid's center. In conclusion, resolving cooled plasmoids on a finite mesh is numerically challenging.

\section{FMS wave injection with limited upstream flux supply}
\label{sec:appendixD}

\begin{figure}
  \centering
  \includegraphics[width=0.5\textwidth]{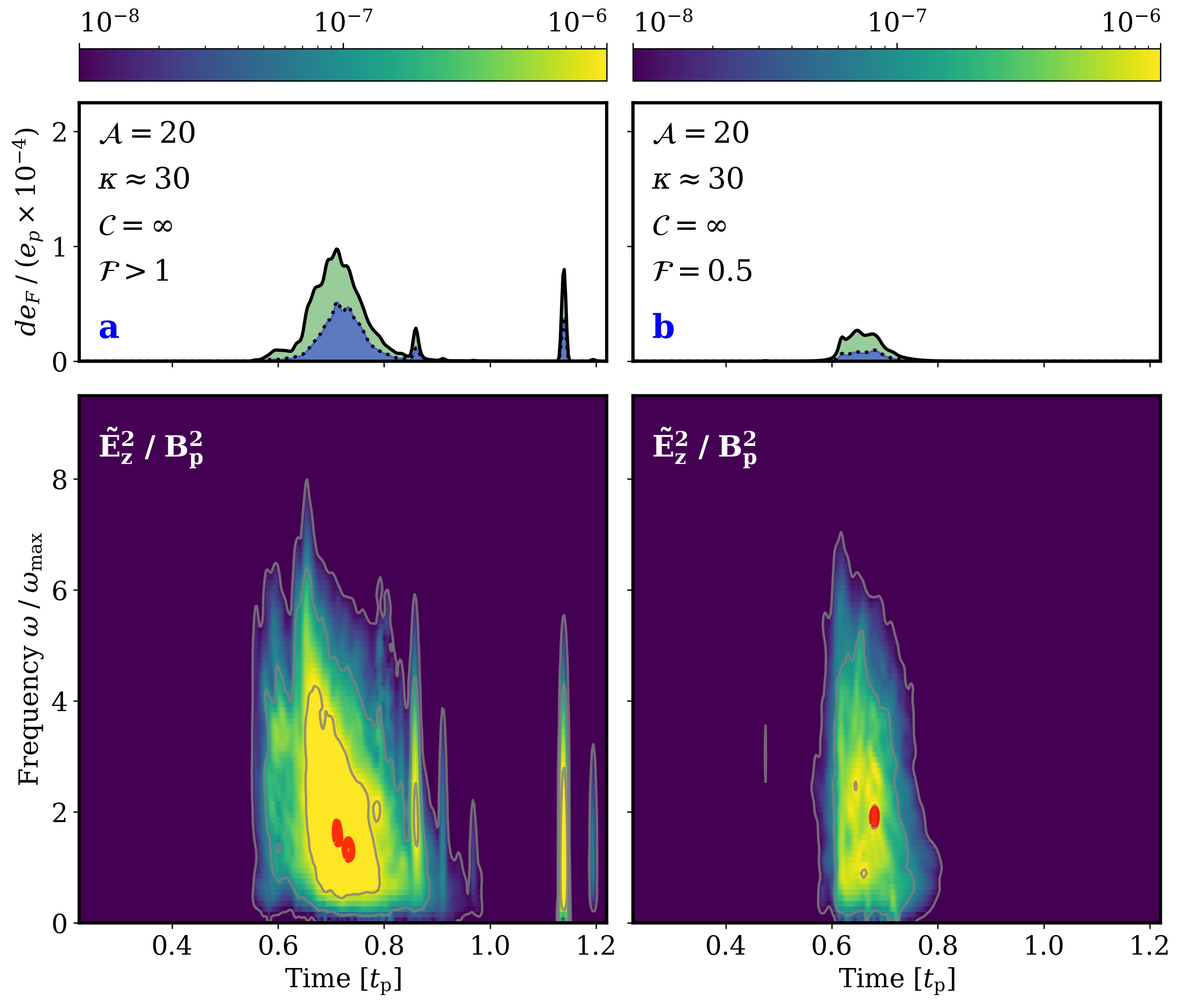}
  \vspace{-5pt}
  \caption{Comparison of dynamic spectra for a varying supply of upstream flux. Panel (a) shows an $\mathcal{A}=20$ reference case for a system of infinite length ($\mathcal{F}>1$, see Figure~\ref{fig:dynamicspectra}). Panel (b) shows an adaptation of the standard setup to a finite flux reservoir ($\mathcal{F}=0.5$, Appendix~\ref{sec:appendixD}). We provide an animation of this simulation as supplementary material \citep{SupplementaryMediaB}.}
\label{fig:finiteflux}
\end{figure}

To mimic the effects of a finite flux reservoir on the proposed mechanism for reconnection-mediated FMS wave generation, we analyze a variation of the $\mathcal{A}=20$ reference simulations (Section~\ref{sec:amplitudedependence}). During the interaction of the current sheet with the LFP, at a time $t=1.6\times\Delta/ c< t_{\rm i}$, we inject a second current sheet (and field reversal) close to the upstream boundary. The upstream flux available to support reconnection is then enclosed by the two current sheets and, thus, finite. In practice, the adapted simulations create a scenario with $\mathcal{F}\approx 0.5$ (using $\beta_{\rm rec}\approx 0.2$ as measured in Section~\ref{sec:amplitudedependence} and $t_i$ scaling as derived in Appendix~\ref{sec:appendixA1}). Figure~\ref{fig:finiteflux} shows the dynamic spectra and intensity of the FMS waves injected for the finite flux experiment. As expected, reconnection stops operating well before $t_i$. Thus, the course of FMS wave injection is different compared to the case of the infinite system, especially during the phase of mergers of largest plasmoids. However, the dynamic spectrum and signal duration are only mildly affected, while the total intensity significantly decreases, in line with the expectations discussed in Section~\ref{sec:dicsussion}.

\end{document}